\documentclass[numsec,webpdf,modern,medium,namedate]{oup-authoring-template}

\onecolumn

\theoremstyle{thmstyleone}%
\newtheorem{coro}{Corollary}[section]
\newtheorem{thm}{Theorem}[section]

\newtheorem{defi}{Definition}[section]
\newtheorem{prop}{Proposition}[section]

\theoremstyle{thmstyletwo}%
\theoremstyle{thmstylethree}%

\usepackage[doublespacing]{setspace}
\usepackage[fontsize=12pt]{fontsize}
\usepackage{comment}

\usepackage{paralist}

\usepackage{algorithm}  
\usepackage{algpseudocode} 
\usepackage{dsfont}
\usepackage{graphicx}%插入图片
\usepackage{float}%固定图片的位置
\graphicspath{{figures/}}%文章所用图片在当前目录下的 figures目录
\usepackage{color}
\usepackage{dsfont}
\usepackage{epstopdf}
\usepackage[hang]{subfigure}%多个图片共有
\usepackage{setspace}
\usepackage{natbib} %控制参考文献格式的代码
\usepackage{mathrsfs}% 不带花体小写字母
\usepackage{cases}
\usepackage{bm}
\usepackage{systeme}

\usepackage{ulem}
\usepackage{xcolor}
\usepackage{textcomp}
\usepackage{amsfonts,amssymb}
\usepackage{multirow}
\usepackage{multicol}
\usepackage{enumerate}
\usepackage{threeparttable}

\usepackage{algorithm}
\usepackage{algpseudocode}

\definecolor{red}{RGB}{139,0,18}
\definecolor{lightred}{RGB}{186,25,31}
\definecolor{blue}{RGB}{0,0,255}
\definecolor{lightblue}{RGB}{69,100,139}
\renewcommand\emph[1]{{\color{red}\itshape #1}}

\newcommand\Rm[1]{\uppercase\expandafter{\romannumeral#1}}
\def\lam{\lambda}

\def\diag{\bm {\mbox{diag}}}

\newcommand\norm[1]{\left\Vert #1 \right\Vert}
\newcommand\abs[1]{\left\vert #1 \right\vert}

\newlength{\algowidth}

\newcommand{\com}[1]{\iffalse #1 \fi}
\def\ervm{{\textnormal{m}}}

\def\mR{\mathbb{R}}

\def\calP{\mathcal{P}}
\def\calN{\mathcal{N}}

\def\calS{\mathcal{S}}
\def\calT{\mathcal{T}}

\def\tr{\mbox{tr}}

\def\vec{\mbox{vec}}

\def\cov{\mbox{cov}}

\def\wh{\widehat}
\def\wt{\widetilde}
\def\ba{{\bm\alpha}}

\def\bb{{\bm\beta}}

\def\b{{\bf b}}
\def\A{{\bf A}}
\def\B{{\bf B}}

\def\H{{\bf H}}

\def\m{{\bf m}}
\def\P{{\bf P}}

\def\I{{\bf I}}

\def\M{{\bf M}}
\def\I{{\bf I}}
\def\X{{\bf X}}
\def\x{{\bf x}}

\def\y{{\bf y}}

\def\V{{\bf V}}
\def\bS{{\bf S}}
\def\O{{\bf O}}
\def\v{{\bf v}}

\def\defby{\stackrel{\mbox{\textrm{\tiny def}}}{=}}
\newcommand{\trans}{^{\mbox{\tiny{T}}}}

\newcommand{\bdelta}{\bm \delta}
\newcommand{\beps}{\bm\varepsilon}

\newcommand{\bPsi}{\bm\Psi}
\newcommand{\bpsi}{\bm\psi}
\newcommand{\bfeta}{\bm\eta}
\newcommand{\bSig}{\bm \Sigma}

\numberwithin{equation}{section}

\AtBeginEnvironment{equation}{%
\setlength{\abovedisplayskip}{4pt plus 2pt minus 2pt} 
\setlength{\belowdisplayskip}{2pt plus 2pt minus 2pt}
}

\AtBeginEnvironment{align}{%
\setlength{\abovedisplayskip}{4pt plus 1pt minus 1pt} 
\setlength{\belowdisplayskip}{4pt plus 1pt minus 1pt}
}

\usepackage{xr}
\makeatletter
\g@addto@macro\normalsize{%
  \setlength\abovedisplayskip{3pt plus 1pt minus 1pt}
  \setlength\belowdisplayskip{3pt plus 1pt minus 1pt}
  \setlength\abovedisplayshortskip{2pt plus 1pt minus 1pt}
  \setlength\belowdisplayshortskip{2pt plus 1pt minus 1pt}
}
\makeatother

\begin{document}

\title[Distributed Prediction under Heterogeneity with Unidentifiable Parameter]{Distributed Prediction under Heterogeneity with Unidentifiable Parameter}

\author[1]{Erbo Li}
\author[1]{Zhaojun Hu}
\author[1]{Ting Wei}
%\author[1]{Yuze Han}
\author[1,2$\ast$]{Yifan Sun}
\author[3]{Liping Zhu}

\authormark{Erbo Li et al.}

\address[1]{\orgdiv{Center for Applied Statistics}, \orgname{School of Statistics, Renmin University of China}}
% {\orgdiv{Center for Applied Statistics}, \orgname{School of Statistics, Renmin University of China}, \orgaddress{\state{Beijing}, \country{China}}}
% \address[1]{\orgdiv{Center for Applied Statistics}, \orgname{School of Statistics, Renmin University of China}, \orgaddress{\street{Street}, \postcode{Postcode}, \state{Beijing}, \country{China}}}
\address[2]{\orgdiv{Beijing Advanced Innovation Center for Future Blockchain and Privacy Computing}}
\address[3]{\orgdiv{Center for Applied Statistics}, \orgname{Institute of Statistics and Big Data, Renmin University of China}}

\corresp[$\ast$]{Address for correspondence. Yifan Sun, Center for Applied Statistics, School of Statistics, Renmin University of China, Beijing Advanced Innovation Center for Future Blockchain and Privacy Computing, Beijing, 100872, China. \href{Email:email-id.com}{sunyifan1984@163.com}}

% \abstract{
% Distributed semiparametric dimension reduction frequently encounters computational and statistical bottlenecks driven by objective nonconvexity and network heterogeneity. The nonconvexity primarily stems from the structural non-identifiability of index parameters. To systematically resolve these issues, we propose InvexDR, a communication-efficient distributed estimation framework. By constructing a targeted invex surrogate, our method eliminates suboptimal local minima and guarantees global optimization. To accommodate heterogeneous data, the algorithm executes an adaptive aggregation step, dynamically weighting neighboring estimators based on underlying similarities and local signal-to-noise ratios. Theoretically, we prove that the proposed estimator achieves a minimax optimal, two-phase convergence rate of $O_p(N^{-1/2} + H \wedge n_j^{-1/2})$ over the heterogeneous parameter space, where $N$ and $n_j$ denote the global and local sample sizes, and $H$ bounds the heterogeneity level. Furthermore, we strictly quantify the optimization drift introduced by multi-step local updates, establishing conditions under which the algorithmic sequence achieves statistical parity with the theoretical optimum. Extensive numerical experiments confirm that our approach maintains robust estimation precision while significantly reducing communication rounds across diverse heterogeneous settings.
% }

\abstract{
Predicting a response based on covariates is a fundamental problem in statistics and machine learning. However, profound difficulties arise when the underlying low-dimensional structural parameters are unidentifiable, as typified in dimension reduction contexts. Specifically,estimating these non-identifiable parameters inherently introduces severe nonconvexity. In distributed settings, this difficulty is further compounded by the challenges of data heterogeneity and communication cost.
To overcome these intertwined barriers, we propose a novel distributed semiparametric framework. We formulate an adaptive homogeneity pursuit utilizing a trace-similarity penalty to effectively address data heterogeneity. To resolve the ensuing severe nonconvexity and communication bottlenecks, we introduce an invex relaxation technique coupled with a multi-step local update algorithm, ensuring stable convergence to global optimality with significantly reduced communication overhead. 
Theoretically, we establish a non-asymptotic model-free prediction error bound and prove that our estimator achieves a two-phase minimax optimal convergence rate and an sharper model-free prediction error bound. Furthermore, we provide theoretical guarantees for algorithmic convergence and communication efficiency. 
Extensive simulations and a real-world multi-center medical application validate the superiority of our method.
}
\keywords{ Distributed Algorithm,  Heterogeneity, Nonconvexity, Unidentifiable Parameter}

\maketitle

\section{Introduction} \label{sec:intro}

A central problem in statistics and machine learning is predicting $Y\in\mR$ based on the covariate $\x\in\mR^p$, which typically relies on nonparametric estimation of the regression function $E(Y\mid \x)$ \citep{vcivzek2020robust,caron2022estimating,salibian2023robust}. In high-dimensional settings, identifying a low-dimensional structure $E(Y\mid \x) = E(Y\mid \x\trans\bb)$ for some parameter $\bb \in \mR^{p \times d}$ significantly enhances both estimation accuracy and interpretability. Assuming the semiparametric model $Y = \ervm(\x\trans\bb) + \epsilon$, our goal is to simultaneously recover the structural parameter $\bb$ and the unknown link function $\ervm(\cdot)$ to derive the ultimate prediction with the estimator $(\wh{\ervm}, \wh\bb)$. However, recovering this low-dimensional structure inherently introduces severe parameter non-identifiability. To contextualize this fundamental barrier, we examine three representative statistical models across communications, economics, and medicine \citep{yang2025onebit,lian2021singleindex,alshamrani2025trans}.

\begin{enumerate}
  \setlength{\itemsep}{-2pt} 
  \setlength{\parsep}{0pt}   
  \setlength{\parskip}{0pt}

    \item In 1-bit Compressed Sensing (CS) \citep{fan2023onebit,yang2025onebit}, a central task  is recovering a signal $\bb$ from quantized measurements $Y = \text{sign}(\bb\trans \x + \epsilon)$, where $\ervm(\x\trans\bb) = 1 - 2F_\epsilon(-\x\trans\bb)$, with the cumulative distribution function $F_\epsilon$ for $\epsilon$. Since the sign function's nonlinearity and the unknown noise variance, the signal magnitude is fundamentally lost. Consequently, standard reconstruction methods must restrict recovery to the unit sphere \citep{dai2016noisy}. Corresponding formulation introduces a nonconvex optimization problem, only obtaining the signal's direction \citep{huang2018robust}. 
    
    \item The Transformation Model \citep{gorgens1999trans} unifies key bio-statistical models such as the Box-Cox \citep{Bantis2024trans,alshamrani2025trans} and Tobit models \citep{Amore2021tobit,Liu2023tobit}, where $\Lambda_{0}(Y) = \x\trans \bb + \beps$ with an unspecified strictly increasing transformation $\Lambda_{0}(\cdot)$. Because both the transformation and the error scale are unknown, $\bb$ is structurally non-identifiable. Consequently, classic estimators \citep{chen2002trans,hothorn2014transJRSSB,rahmani2025trans} inevitably collide with this identifiability barrier, severely complicating numerical optimization.
    
    \item Widely adopted in econometrics \citep{horowitz2009SIM}, the Single Index Model (SIM) specifies $E(Y\mid \x) = \ervm(\x\trans\bb)$ with an unknown link $\ervm(\cdot)$ that structurally masks the scale of $\bb$. To secure identifiability, various semiparametric and PAC-Bayesian methods \citep{dudeja2018SIM, alquier2013SIM} artificially restrict the parameter to the unit sphere ($\norm{\bb} = 1$) \citep{feng2013partial,zhang2012dimension}. Imposing this constraint injects severe nonconvexity into the objective function, rendering estimation suboptimal.
\end{enumerate}

These models exemplify unidentifiable parameters, and  all can be framed as special cases of sufficient dimension reduction (SDR). SDR targets the central mean space $\calS_{E(Y\mid\x)}$ \citep{li2018SDR}, where $E(Y\mid \x) = E(Y\mid \x\trans\bb)$. While various approaches estimate this space \citep{li1991SIR,xia2002mave}, they universally confront parameter non-identifiability. 

This inherent non-identifiability precipitates the first fundamental challenge: nonconvexity. To meaningfully quantify parameter similarity, one must rely on projection matrices, $\P(\bb) = \bb(\bb\trans\bb)^{-1}\bb\trans$, when one utilizes metrics like the Frobenius norm or trace inner product \citep{zeng2024SDRjasa,Xia2025DPsdr}. However, the highly nonlinear projection operator $\P(\cdot)$ inevitably injects severe nonconvexity into the optimization. Existing workarounds, such as imposing block identity structures or restricting estimators to specific manifolds \citep{ma2013semiDRjrssb,wenzaiwen2013orth}, either conflict with the true parameter structure or induce optimization difficulties.

Beyond nonconvexity in the centralized setting, distributed computing introduces a second major challenge: communication overhead. Driven by strict privacy constraints and edge device computing capability, modern collaborative training inherently relies on parameter communication. While distributing $N$ samples across $m$ nodes drastically reduces computational complexity from $O(N^2)$ to $O\{(N/m)^2\}$ in semiparametric SDR \citep{zhu2025distributed}, limited network bandwidth renders this communication a critical bottleneck \citep{Jordan2019communication,fan2023communication}, necessitating communication-efficient algorithms \citep{chen2022distributed,zhu2022distributed,cai2020dsir}.

Finally, the proliferation of federated learning (FL) introduces the third pervasive challenge: data heterogeneity. In many distributed scenarios, local nodes often possess distinct structural parameters $\bb$ \citep{nguyen2021FLsurvey}. While transfer learning (TL) and multi-task learning (MTL) effectively leverage Euclidean distances to aggregate identifiable parameters \citep{Li2022TLjrssb,zhang2022mtlsurvey,wang2023TLnonpara}, Euclidean metrics completely fail for non-identifiable parameters. They inherently cannot capture similarities between parameters that share the exact same column space but differ merely in magnitude \citep{wang2025robust,maity2019mtl}.

\subsection{Our Contributions}

Estimating non-identifiable parameters in distributed settings intertwines three formidable challenges: severe nonconvexity, data heterogeneity, and communication overhead. Utilizing trace inner products to aggregate heterogeneous signals structurally exacerbates nonconvexity. However, existing workarounds fall short: convex relaxations \citep{Gu2022angleTL} fail to adaptively capture parameter homogeneity under non-identifiability, whereas directly estimating the expanded  central space span \citep{Xu2023hSDR} incurs prohibitive communication costs without yielding node-specific estimates. Therefore, developing a communication-efficient distributed framework for heterogeneous, non-identifiable parameters remains a critical open problem.

To bridge this gap, we propose a semiparametric distributed framework that systematically resolves these three challenges. First, we tackle data heterogeneity by integrating a trace-similarity penalty into a Newton-Raphson least squares objective, directly achieving adaptive homogeneity pursuit across nodes. Because this trace penalty structurally exacerbates nonconvexity, we introduce an invex relaxation \citep{Barik2022invexmlr, Barik2023InvexPF}. This transformation ensures that the Karush--Kuhn--Tucker (KKT) conditions guarantee global optimality, creating a highly stable landscape for first-order optimization \citep{Hanson1981invex, Barik2021invex}. Finally, building strictly upon this invex foundation, we deploy a multi-step local update strategy to substantially reduce communication overhead without risking algorithmic divergence.

Theoretically, we rigorously establish a comprehensive foundation for both statistical optimality and algorithmic efficiency. Statistically, our method improves the estimation rate from the local $O_p(n^{-1/2})$ to the distributed $O_p(N^{-1/2} + H \land n^{-1/2})$, alongside a corresponding model-free prediction upper bound. We prove this rate achieves minimax optimality over an information set characterized by trace similarity. Notably, our minimax lower bound matches the order of heterogeneous frameworks explicitly designed for identifiable parameters \citep{Li2022TLjrssb, Li2023TLjasa}.  Algorithmically, we guarantee stable convergence to the minimax estimator from arbitrary initializations, strictly bounding the communication cost to $O(pK^{1/2})$, where $p$ is the dimension and $K$ the local iteration steps.

Empirically, extensive simulations validate our theoretical convergence rates and demonstrate consistent superiority over baseline methods. Furthermore, a real-world multi-center application predicting the length of stay for mildly comatose ICU patients \citep{sheikhalishahi2020benchmarking} confirms the practical utility and robust predictive accuracy of our framework in authentic distributed healthcare settings.

Our main contributions are summarized as follows:
\begin{enumerate}
  \setlength{\itemsep}{-2pt} 
  \setlength{\parsep}{0pt}   
  \setlength{\parskip}{0pt}

    \item We propose a robust distributed semiparametric framework to tackle data heterogeneity. By integrating a Newton--Raphson least squares formulation with a trace-similarity penalty, our method achieves adaptive homogeneity pursuit for estimating non-identifiable parameters across diverse nodes.
    
    \item We resolve the severe nonconvexity induced by the trace penalty via an invex relaxation. This transformation preserves the formulation's adaptive properties while guaranteeing stable convergence to a global optimum from arbitrary initializations.
    
    \item We establish comprehensive theoretical guarantees for both statistical optimality and communication efficiency. Statistically, our estimator achieves a minimax-optimal rate of $O_p(N^{-1/2} + H\land n^{-1/2})$, ensuring superior model-free prediction. Algorithmically, our multi-step update strategy strictly bounds the communication overhead to $O(mK^{1/2})$.
\end{enumerate}

The remainder of this paper is organized as follows. Section~2 introduces the distributed methodology and multi-step algorithm. Section~3 establishes the non-asymptotic minimax optimality alongside algorithmic and communication guarantees, while Sections~4 and~5 evaluate empirical performance via simulations and a real-data application. Section~6 concludes, with all technical proofs deferred to the supplementary material.

\section{Methodology}
% \section{Formualtion Development}

This section details the proposed methodology, sequentially presenting the statistical formulation, the optimization design, and the distributed algorithm. First, we propose a framework to adaptively aggregate heterogeneous, non-identifiable parameters based on a semiparametric least squares method. Second, to overcome the resulting nonconvexity, we introduce an invex relaxation technique. Finally, leveraging this stable optimization foundation, we design a communication-efficient distributed algorithm. This progression can systematically resolve the challenges of data heterogeneity, nonconvexity, and communication costs.

The following notations are used throughout this paper. Generic constants are denoted by $C, c, C_0, c_0, \ldots$, whose values may vary from line to line. For a vector $\ba \in \mathbb{R}^p$, $\norm{\ba}_q$ denotes its $\ell_q$-norm. For a matrix $\A \in \mathbb{R}^{p \times d}$, we let $\gamma_{\max}(\A)$, $\gamma_{\min}(\A)$, and $\gamma_l(\A)$ denote its maximum, minimum, and $l$-th largest eigenvalues, respectively, and $\tr(\A)$ denote its trace. Its spectral and Frobenius norms are denoted by $\norm{\A}_2$ and $\norm{\A}_F$, respectively. The operator $\vec(\A)$ vectorizes $\A$ by stacking its columns, with $\vec^{-1}(\cdot)$ serving as its column-major reshaping inverse. Let $\I_p$ be the $p$-dimensional identity matrix, and $\mathbf{P}(\A) \defby \A(\A\trans \A)^{-1} \A\trans$ be the orthogonal projection matrix onto the column space of $\A$. For random sequences $X_n$ and $Y_n$, we write $X_n = O_p(Y_n)$ if, for any $\epsilon > 0$, there exists a constant $C > 0$ such that $\Pr(|X_n| \leq C|Y_n|) \geq 1 - \epsilon$ holds for all sufficiently large $n$. For two non-negative sequences $a_n$ and $b_n$, we write $a_n \lesssim b_n$ and $a_n \gtrsim b_n$ to denote that $a_n \leq C b_n$ and $a_n \geq C b_n$, respectively, for some constant $C > 0$ and all sufficiently large $n$.

\subsection{A Regularized Optimization Formulation with Trace Penalty}

In this section, we introduce a regularized optimization framework combining a semiparametric least squares loss with a trace-based penalty to achieve adaptive homogeneity pursuit across heterogeneous, non-identifiable parameters.

\subsubsection{Semiparametric Least Squares Formulation}
To formalize this approach, we first revisit the classical semiparametric least squares
framework. Let $\calS_{E(Y\mid\x)}$ be the central mean subspace spanned by a basis matrix $\bb$. The conditional mean model satisfies
\begin{equation}
    Y = \ervm(\bb\trans\x) + \epsilon, 
    \label{eq:LowStruct}
\end{equation}
where $\ervm(\bb\trans\x) \defby E(Y\mid\bb\trans\x)$ is an unspecified smooth link function, and $E(E\mid\x) = 0$. Following \cite{ma2013semiDRjrssb,zhu2025distributed}, we estimate $\bb$ by setting the inverse conditional variance weight to unity. This specification substantially reduces computational costs while rigorously preserving estimator consistency.

We solve the unweighted population-level estimating equation:
\begin{equation} \label{eq:form:ee}
    E\left[\bS\left(\x, Y, \bb\right)\right] = \mathbf{0},
\end{equation}
where $\mathbf{S}\left\{\x, Y, \ba\right\} \defby \{Y - \ervm(\x\trans\ba)\} \wt\x(\ba)$, and the adjusted covariate vector $\wt\x(\ba) \in \mathbb{R}^{pd}$ is defined as
\begin{equation}
\wt\x(\ba) \defby \vec\Big\{ 
    \big[ 
        \x - E\left( \x \mid \x\trans \ba \right)
    \big] 
    \mathbf{m}_{1}\trans \left( \x\trans \ba \right)
\Big\},
\end{equation}
with $\mathbf{m}_1(\x\trans\ba) \defby \partial \ervm(\x\trans\ba) / \partial (\x\trans\ba)$ denoting the link function gradient.

Given an initial value $\bb^{(0)}$, Equation \eqref{eq:form:ee} is typically solved via the Newton--Raphson update:
\begin{equation} \label{eq:NRite}
    \vec(\bb^{(t+1)}) \defby \vec(\bb^{(t)}) + \{\H(\bb^{(t)})\}^{-1}E\left[\mathbf{S}\left\{\x, Y, \bb^{(t)}\right\}\right],
\end{equation}
where $\H(\ba) \defby E\big[\{\wt\x(\ba)\}\{\wt\x(\ba)\}\trans\big]$. Following \cite{ma2013semiDRjrssb}, we recast this step as an equivalent iterative least squares problem at the $(t+1)$-th iteration:
\begin{equation} \label{model:NRLS}
    \vec(\bb^{(t+1)}) = \arg\min_{\ba} E\left[\big\{\wt{Y}(\bb^{(t)}) - \wt\x(\bb^{(t)})\trans\vec(\ba)\big\}^2 \right],
\end{equation}
where the pseudo-response is $\wt{Y}(\ba) \defby \{\wt\x(\ba)\}\trans\vec(\ba) + \{Y - \ervm(\x\trans\ba)\}$.

\subsubsection{Heterogeneity Aggregation via Trace Penalty}
Building upon this population-level framework, we now introduce the Heterogeneity Aggregation method and the corresponding heterogeneous distributed setting.
Consider a distributed system with $m$ distinct nodes, where node $j \in \{1,\dots,m\}$ holds $n_j$ independent observations $\{(\x_{i,j}, Y_{i,j})\}_{i=1}^{n_j}$. To account for structural heterogeneity, each node possesses a distinct local central mean subspace spanned by $\bb_j^* \in \mathbb{R}^{p\times d}$. Our goal is to jointly estimate the global parameter matrix $\bb^* = \big(\bb_1^*,\dots,\bb_m^*\big) \in \mathbb{R}^{p\times md}$.

We nonparametrically estimate the unknown local components on each node. For a local parameter $\ba_j$, local linear regression yields the link function $\wh{\ervm}_j(\x_{i,j}\trans\ba_j) = \wh{b}_{i,j}$ and its gradient $\wh{\m}_{1,j}(\x_{i,j}\trans\ba_j) = \wh{\b}_{i,j}$ by solving
\begin{equation} \label{eq:K1}
    \small
    (\wh{b}_{i,j},\wh{\b}_{i,j}) \defby \arg\min\limits_{{b}_{i,j},{\b}_{i,j}}
    \sum_{k \neq i}^{n_j}
    \big\{ Y_{k,j} - b_{i,j} - ( \x_{k,j} - \x_{i,j} )\trans \ba_j \mathbf{b}_{i,j} \big\}^{2}
    K_{h_{1}} \big\{ ( \x_{k,j} - \x_{i,j} )\trans \ba_j \big\}.
\end{equation}
Concurrently, Nadaraya--Watson smoothing estimates the conditional covariate expectation:
\begin{equation} \label{eq:K2}
     \wh{E}_{j} \left\{ \x_{j} \mid \x_{i,j}\trans \ba_j \right\} \defby 
    \frac{\sum_{k \neq i}^{n_j} K_{h_{2}} \big\{ ( \x_{k,j} - \x_{i,j} )\trans \ba_j \big\} \x_{k,j} }
    {\sum_{k \neq i}^{n_j} K_{h_{2}} \big\{ ( \x_{k,j} - \x_{i,j} )\trans \ba_j \big\}}.
\end{equation}
These local estimators define the empirical adjusted covariate vector $\wh{\x}_{i,j}(\ba_j) \in \mathbb{R}^{pd}$ and pseudo-response $\wh{Y}_{i,j}(\ba_j) \in \mathbb{R}$:
\begin{align}
        \wh{\x}_{i,j}(\ba_j) &\defby \vec\Big[ \big\{ \x_{i,j} - \widehat{E}_{j}( \x_{j} \mid \x_{i,j}\trans \ba_j ) \big\} \wh{\m}_{1,j}\trans ( \x_{i,j}\trans \ba_j ) \Big], \label{eq:hatx} \\ 
        \wh{Y}_{i,j}(\ba_j) &\defby \wh{\x}_{i,j}(\ba_j)\trans\vec(\ba_j) + \big\{ Y_{i,j} - \wh{\ervm}_{j}( \x_{i,j}\trans \ba_j ) \big\}. \label{eq:hatY}
\end{align}
Accordingly, the sample estimating function and the associated negative gradient matrix for node $j$ are given by
\begin{equation} \label{eq:form:ee:emp}
    \wh{E}_j \left\{ \bS\left(\x, Y, \bb\right) \right\} = \{Y - m_j(\x_j\trans\ba)\} \wh\x_j(\ba),
\end{equation}
\begin{equation}
    \wh{\H}_j(\ba) \defby n_j^{-1} \sum_{i=1}^{n_j}\big[\{\wh\x_{i,j}(\ba)\}\{\wh\x_{i,j}(\ba)\}\trans\big].
\end{equation}

While the local empirical loss efficiently yields individual estimates, the inherent non-identifiability of these low-dimensional parameters poses a critical challenge. As discussed in Section~\ref{sec:intro}, parameter similarity can only be meaningfully quantified through their respective column spaces. This intrinsic geometric constraint renders conventional Euclidean distances inadequate. Consequently, standard distributed aggregation strategies relying on Euclidean metrics are fundamentally inapplicable, such as $\ell_2$-norm regularization in FedProx \citep{yuan2022convergence} or naive fusion penalties \citep{hand2016convex, ma2017concave}.

To overcome this limitation and enable a meaningful pursuit of homogeneity, we formulate the global parameter aggregation at the $(t+1)$-th iteration as a nonconvex optimization problem. The global objective synergistically integrates the local empirical least squares loss with a trace-based subspace penalty:
\begin{align} \label{prob:original}
    \wh\bb^{(t+1)} &= \arg\min_{\ba} F(\ba;\lam,\wh\bb^{(t)}), \quad F(\ba;\lam,\wh\bb^{(t)}) \defby \sum_{j=1}^{m} \left\{ L_j(\ba_j;\wh\bb^{(t)}) + \frac{\lam}{m} P_j(\ba_j;\wh\bb^{(t)}) \right\},
\end{align}
where $\ba = (\ba_1,\dots,\ba_m) \in \mathbb{R}^{p\times md}$. Here, the local empirical loss evaluates the iterative least squares objective \eqref{model:NRLS} using the node-specific empirical components:
\begin{equation} \label{prob:original:loss}
    L_j(\ba_j;\bb^{(t)}) \defby \sum_{i=1}^{n_j} \big\{ \wh{Y}_{i,j}(\bb_j^{(t)}) - \wh{\x}_{i,j}\trans(\bb_j^{(t)})\vec(\ba_j) \big\}^2.
\end{equation}
Meanwhile, the trace penalty $P_j$, designed to structurally enforce subspace similarity across nodes, is defined as
\begin{equation} \label{prob:original:penalty}
    P_j(\ba_j;\bb^{(t)}) \defby - \frac{1}{d} \sum_{l\ne j} \tr\big\{\P(\ba_j)\P(\bb_l^{(t)})\big\},
\end{equation}
where $\P(\A) \defby \A(\A\trans\A)^{-1}\A\trans$ denotes the orthogonal projection matrix onto the column space of $\A$.

To elucidate the geometric mechanism of this penalty, let $\V(\ba_j) = (\v_1(\ba_j),\dots,\v_d(\ba_j)) \in \mathbb{R}^{p \times d}$ denote an orthonormal basis matrix for the column space of $\ba_j$, yielding $\P(\ba_j) = \V(\ba_j)\V(\ba_j)\trans$. It follows that
\begin{equation}
    \tr\big\{\P(\ba_j)\P(\bb_l^{(t)})\big\} = \sum_{l=1}^{d}\sum_{l'=1}^{d} \langle\v_l(\ba_j),\v_{l'}(\bb_l^{(t)})\rangle^2,
\end{equation}
where $ \langle\cdot,\cdot\rangle$ is the angle between the corresponding vectors. This geometric identity guarantees that $-1 \le P_j(\ba_j;\bb^{(t)}) \le 0$. By minimizing the overall objective $F$, the trace penalty actively maximizes these squared inner products. Consequently, it adaptively aligns the local subspaces toward the homogeneous structure, ultimately yielding a more statistically efficient estimator.

\subsection{Invex Relaxation}

Although subsequent theoretical results demonstrate that the trace penalty effectively drives adaptive heterogeneous aggregation, it inherently injects non-convexity into the objective function. Conventional convex relaxations can restore solvability, but they often discard the intrinsic geometric information of the original parameter space \citep{Gu2022angleTL}. 

To bypass this computational barrier while strictly preserving structural integrity, we propose a specialized invex relaxation. By reformulating the problem into an invex function \citep{Barik2021invex,Hanson1981invex}, we secure a powerful computational guarantee: the Karush-Kuhn-Tucker (KKT) conditions remain both necessary and sufficient for global optimality. 

Specifically, we augment the parameter dimension and propose the following relaxed objective function for the original problem \eqref{prob:original}:
\begin{equation}
    G(\A;\lam,\bb^{(t)}) \defby \sum_{j=1}^{m} G_j(\A;\lam,\bb^{(t)}), \quad G_j(\A;\lam,\bb^{(t)}) \defby J_j(\A) + (\lambda/md)R_j(\A;\bb^{(t)}) \label{prob:invex:obj}
\end{equation}
where $\A = (\A_1, \dots, \A_m) \in \mathbb{R}^{(p+d) \times md}$. The local empirical loss and the trace penalty are respectively recast as
\begin{align}
    J_j(\A_j;\bb^{(t)}) &= {n_j}^{-1} \sum_{i=1}^{n_j} \tr\left\{\bS_{i,j}\trans(\bb^{(t)}) \A_j\right\}^2, \label{prob:invex:loss}\\
    R_j(\A_j;\bb^{(t)}) &= (md)^{-1} \sum_{l \ne j} \tr\left\{\wt\P(\A_j) \wt\P(\A_l)\right\}, \label{prob:invex:penalty}
\end{align}
where
\begin{equation}
        \wt\P(\B) \defby \M_2 \B \left\{(\B - \M_1)\trans (\B - \M_1)\right\} \B\trans \M_2, \label{prob:invex:proj} 
\end{equation}
and the involved constant matrices and empirical components are defined as
\begin{equation}
    \M_1 \defby \begin{bmatrix} \O_{p \times d} \\ \I_d \end{bmatrix}, \quad
    \M_2 \defby \begin{bmatrix} \I_p & \\ & \O_d \end{bmatrix}, \quad
    \bS_{i,j}(\bb^{(t)}) \defby \begin{bmatrix} \wh{\x}_{i,j}(\bb^{(t)}) \\ d^{-1} \wh{Y}_{i,j}(\bb^{(t)}) \I_d \end{bmatrix}.
\end{equation}
The relaxed optimization problem is consequently formulated as
\begin{equation}
    \widehat{\A} = \arg\min_{\A} \; G(\A;\lam,\bb^{(t)})  \quad \text{subject to} \quad \M_1\trans \A_j = \I_d, \quad j =1,\dots,m,
    \label{prob:invex}
\end{equation}
where the constraint is equivalent to $(\A_j)_{(p+1:p+d)\times(1:d)} = \I_d$. The geometric intuition behind this relaxation is straightforward: any feasible augmented matrix naturally partitions as $\A_j = [\bb_j\trans, \I_d]\trans$ for some $\bb_j \in \mathbb{R}^{p \times d}$. Under this partition, the relaxed components exactly recover the original loss and penalty, yielding $J_j(\A_j) = L_j(\ba_j)$ and $R_j(\A_j;\bb^{(t)}) = P_j(\ba_j;\bb^{(t)})$. Proposition~\ref{prop:minimum} formally establishes this one-to-one equivalence, justifying the substitution of the relaxed problem as a valid surrogate.

\begin{prop}\label{prop:minimum}
Let $\wh\bb$ be a global minimum of the original problem \eqref{prob:original}. Then the augmented matrix $\wh\A$ defined by $\wh{\A}_j = [\wh\bb_j\trans, \I_d]\trans$ is a global minimum of the relaxed problem \eqref{prob:invex}. Conversely, if $\wh\A$ is a global minimum of \eqref{prob:invex} such that each block $\wh{\A}_j$ takes the form $[\wh\bb_j\trans, \I_d]\trans$, then $\wh\bb = (\wh\bb_1,\dots,\wh\bb_m)$ is a global minimum of the original problem \eqref{prob:original}.
\end{prop}

To establish the computational tractability of \eqref{prob:invex}, we demonstrate its invexity. An invex function is a broader class of non-convex functions that fundamentally retains the most desirable geometric property of convexity: every stationary point (i.e., a point satisfying the KKT conditions) is automatically a global minimum. For clarity, we first recall the standard definition of invexity.

\begin{defi}[Invex function] \label{Def:Invexity}
Let $\phi(t)$ be a differentiable function defined on a set $\mathcal{T}$. Let $\bfeta: \mathcal{T} \times \mathcal{T} \rightarrow \mathcal{T}$ be a vector-valued function. The function $\phi(t)$ is said to be $\bfeta$-invex if, for all $t_1, t_2 \in \mathcal{T}$, it satisfies
\begin{equation}
    \phi(t_1) - \phi(t_2) \ge \bfeta(t_1, t_2)\trans\nabla\phi(t_2).
\end{equation}
\end{defi}

Invexity defines a special class of non-convex functions that encompasses convexity as a special case (where $\bfeta(t_1,t_2) = t_1 - t_2$). Although the relaxed problem \eqref{prob:invex} remains nonconvex, its invexity can guarantee that all stationary points are globally optimal. Furthermore, because the Polyak-Lojasiewicz (PL) inequality, which guarantees a linear convergence rate, is inherently linked to invexity \citep{karimi2016linear}, gradient-based algorithms applied to this problem are expected to achieve global convergence. 
% Therefore, compared to standard convex relaxations, our invex relaxation achieves reliable computational stability while preserving substantially more of the original structural properties.
Consequently, our invex relaxation fundamentally outperforms standard convex relaxations by simultaneously guaranteeing computational stability and substantially preserving the original structural integrity.
Proposition~\ref{prop:Invexity} rigorously establishes that our relaxed objective function is indeed invex.
% Since invexity encompasses convexity as a special case (where $\bfeta(t_1,t_2) = t_1 - t_2$). Although the relaxed problem \eqref{prob:invex} remains nonconvex, its invexity can guarantee that all stationary points are globally optimal. Furthermore, because the Polyak-Lojasiewicz (PL) inequality, which guarantees a linear convergence rate, is inherently linked to invexity \citep{karimi2016linear}, gradient-based algorithms applied to this problem are expected to achieve global convergence.

% an invex relaxation preserves far more structural properties of the original non-convex problem than a convex relaxation. 

\begin{prop} \label{prop:Invexity}
For any $\A$ within the feasible domain $\mathcal{T}_{\A}$, the problem \eqref{prob:invex} is $\bfeta$-invex concerning $\A$, where $\bfeta (\A,\wt\A) = \diag\left\{\bfeta_1(\A_1,\wt{\A}_1),\dots,\bfeta_m(\A_m,\wt{\A}_m)\right\}$. Specifically,
    \begin{equation*}
    \bfeta_j\left(\A_j,\wt\A_j\right) = \tau_j(\A_j, \wt\A_j) \A_j - \frac{1}{2} \A_j,
    \end{equation*}
    with the scalar $\tau_j$ defined as
    \begin{equation}
    \tau_j = \frac{1}{2} \sum_{l \ne j} \tr\left\{\wt\P(\A_j) \wt\P(\A_l) \wt\P(\A_j)\right\}^{-1} R_j(\wt\A_j;\bb^{(t)}).
    \end{equation}
\end{prop}

In summary, this invex relaxation expands the solution space to safely bypass the original non-convex barriers while perfectly preserving the global minima. Computationally, it guarantees that any gradient-based optimization algorithm satisfying the KKT conditions will definitively converge to the global optimum. This powerful property ensures stable, initialization-independent parameter estimation, thereby providing a rigorous theoretical foundation for our distributed algorithmic design.

% \subsection{Communication-Efficient Distributed Algorithm}

% To optimize the proposed objective function, we develop Invex Dimensional Reduction (InvexDR), a communication-efficient distributed algorithm governed by a bi-level iterative architecture. The outer loop executes a nonparametric Newton-Raphson procedure to sequentially update the local link functions and their gradients. Within this, an inner loop performs projected gradient descent on the invex relaxation to adaptively aggregate structural heterogeneity. Crucially, to strictly control the communication cost, our algorithm executes multiple local updates with gradient descent steps before the aggregation round. 

% To ensure valid evaluations of the nonparametric components and tangent spaces from the outset, we initialize the algorithmic workflow with a reasonably robust estimator. As established in our subsequent theoretical analysis, the algorithmic convergence relies on remarkably mild conditions regarding this initial value. For practical implementation, we initialize the procedure using the MAVE method \citep{xia2002adaptive}. Given its intrinsic structural compatibility with the semiparametric least squares framework, MAVE efficiently provides the robust initial estimators $\bb_j^{(0)}$ required to launch the outer-loop iterations.

\subsection{Communication-Efficient Distributed Algorithm}

To optimize the proposed objective function, we develop Invex Dimensional Reduction (InvexDR), a communication-efficient distributed algorithm featuring a bi-level architecture. The outer loop employs a nonparametric Newton--Raphson procedure to update local link functions and their gradients, while the inner loop performs projected gradient descent on the invex relaxation to adaptively aggregate structural parameters. To rigorously control communication overhead, the inner loop executes multiple local gradient updates before each global aggregation round.

While valid initial evaluations of the nonparametric components in the outer loop require a robust initialization, our theoretical analysis establishes that convergence relies on remarkably mild conditions. Practically, we initialize the procedure using MAVE \citep{xia2002adaptive}. Its inherent structural compatibility with the semiparametric least squares framework efficiently yields reliable initial estimators $\bb_j^{(0)}$ to launch the workflow.

At the $t$-th global communication round, each local node $j$ first performs the outer-loop nonparametric treatment. Specifically, based on the current estimate $\bb_j^{(t)}$, node $j$ estimates the link function $m_j(\x_j\trans\bb_j)$, its gradient $\m_{1,j}(\x_j\trans\bb_j)$, the conditional expectation $E\{\x_j\mid\x_j\trans\bb_j\}$ and the empirical adjusted sample $(\wh\x_{i,j}(\bb_j),\wh{Y}_{i,j}(\bb_j))$ using equations \eqref{eq:K1} -- \eqref{eq:hatY}, with $\bb_j = \bb_j^{(t)}$.

Following the nonparametric update, the algorithm initiates the inner loop with aggregated parameters for distinctive parameter updates on each node. Instead of exchanging raw parameters, nodes share their structural information via an aggregated projection matrix. For node $j$, the central server constructs and broadcasts the structural proxy:
\begin{align}
    \bpsi_j^{(t,k)} = \sum_{l\ne j} \P \left(  \bb_l^{(t,k)}  \right),
    \label{eq:algo:compara}
\end{align}
where $\bb_l^{(t,k)} = \left( \A_j^{(t,k)} \right)_{1:p,1:d}$ is obtained from the relaxation parameter $\A_j^{(t,k)}$.
Utilizing $\bpsi_j^{(t)}$, node $j$ performs $K$ local iterations of projected gradient descent to minimize the invex objective. To foster communication efficiency, global synchronization of $\bpsi_j$ only occurs every $R$ local steps. Let $\eta$ denote the learning rate. A single local gradient update takes the form:
\begin{equation}
    \A_j^{(t,k+1)} = \calP_{\calT_\A}
    \left[
        \A_j^{(t,k)} - 2\eta\nabla\wt{G}_j\left(\A_j^{(t,k)}; \bpsi_j^{(t,k)}\right)
    \right],
\end{equation}
Here, we calculate the local gradient based on $\bpsi_j^{(t,k)}$ as follows: 
%使得 $\nabla\wt{G}_j\left(\A_j^{(t,k)}; \bpsi_j^{(t,k)}\right) = \nabla{G}_j\left(\A_j^{(t,k)}; \bpsi_j^{(t,k)}\right)$
\begin{equation}
    %\small
    \nabla\wt{G}_j\left(\A_j^{(t,k)}; \lam,\bb^{(t)},\bpsi_j^{(t,k)}\right) = \nabla{J}_j(\A_j^{(t,k)};\bb^{(t)})
    -(\lam/md)\nabla\wt{R}_j(\A_j^{(t,k)};\bpsi_j^{(t,k)})
    \label{eq:algo:grad}
\end{equation}
where the gradient of the relaxed loss and penalty is given by
\begin{align}
    \nabla{J}_j(\A_j^{(t,k)};\bb^{(t)}) &= 2{n_j}^{-1} \sum_{i=1}^{n_j} \tr\left\{\bS_{i,j}\trans (\bb^{(t)})\A_j\right\}\bS_j(\bb^{(t)})
    \label{eq:algo:grad:J}
    \\
    \nabla\wt{R}_j(\A_j^{(t,k)};\bpsi_j^{(t,k)}) &= \nabla_{\A_j}\tr\left[\wt\P\left\{\bPsi(\bpsi_j^{(t,k)}) \right\}\wt\P(\A_j)\right],
    \label{eq:algo:grad:R}
\end{align}
where $\bPsi(\bpsi_j^{(t,k)})  = [(\bpsi_j^{(t,k)})\trans,\I_d]\trans$.

Here, the operator $\calP_{\calT_\A}$ projects the updated matrix back onto the feasible domain $\calT_\A \defby \left\{ \A \mid  \M_1\trans\A_j = \I_d \right\}$, defined via the Euclidean projection:
\begin{equation}
    \calP_{\calT_\A} \wt\A_j = \arg\min_{ \M_1\trans\A_j = \I_d} \norm{\A_j -\wt\A_j}_F.
\end{equation}
Upon completing the $K$ inner iterations, the updated parameter is simply extracted as $\bb_j^{(t+1)} = \left( \A_j^{\{t+1\}}   \right)_{1:p,1:d}$. %Node $j$ then transmits $\bb_j^{(t+1)}$ back to the server to conclude the current communication round. 
The complete procedure is summarized in Algorithm~\ref{Algo:invexDR} in Appendix~A.1, as Figure \ref{Figure:AlgoFlow} illustrates.

\begin{figure}[!htbp]
    % \flushleft %左对齐
    \centering
        \includegraphics[width =1\textwidth]{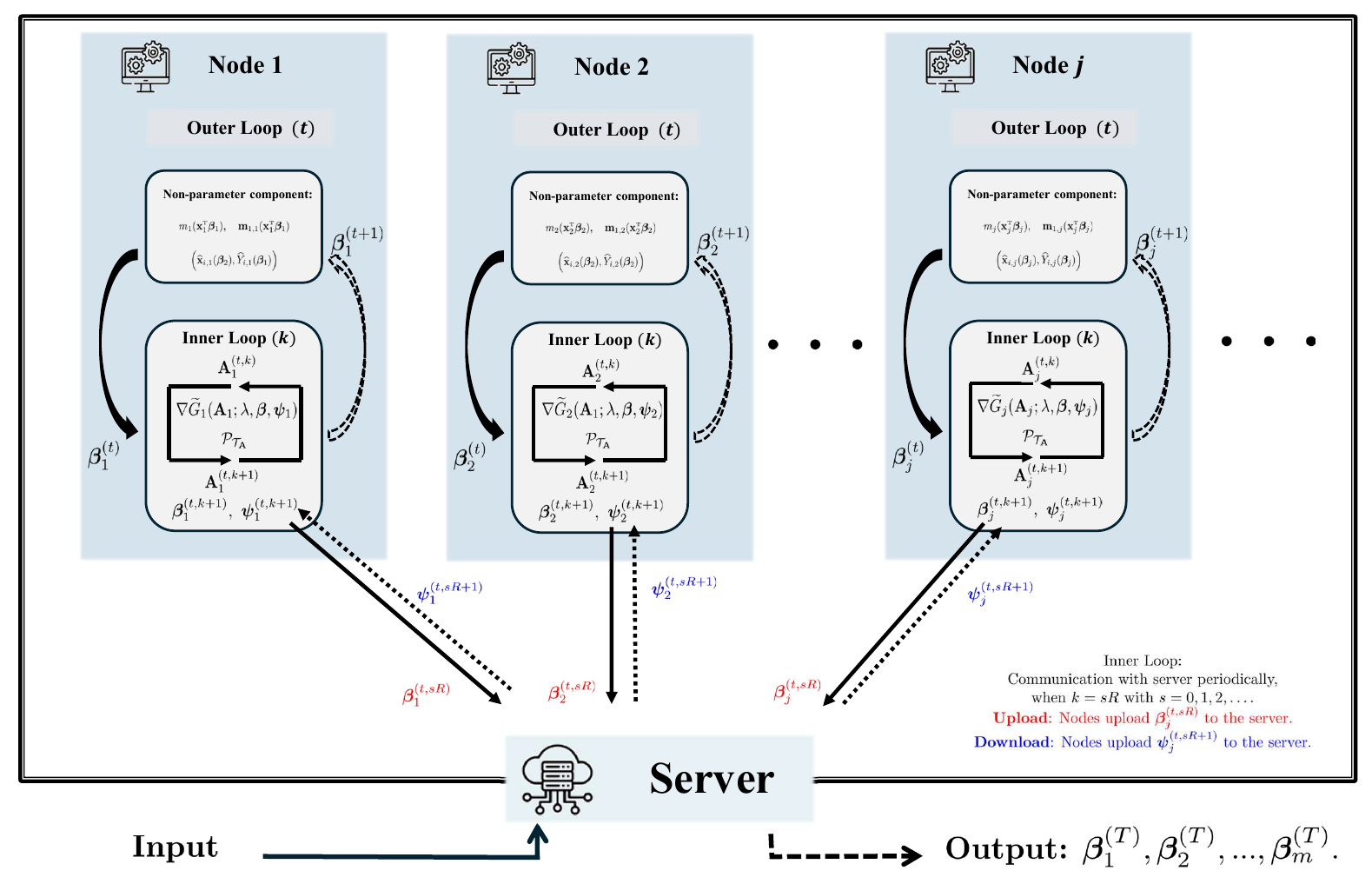}
    \caption{ The workflow of the communication-efficient distributed algorithm with a bi-level loop and local multi-step updates.}
    \label{Figure:AlgoFlow}
\end{figure}

Algorithm~\ref{Algo:invexDR} operates through a bi-level alternating structure. The outer loop estimates the nonparametric components, requiring only a mild initialization that the classical MAVE method readily satisfies. Conditioned on these nonparametric estimates, the inner loop refines the parameter estimation. By leveraging the invex relaxation, this inner optimization is robust to initial values and strictly guarantees convergence to the global optimum. Through the iterative refinement of both nonparametric components and parameter estimation, our subsequent theoretical analysis ensures that the overall algorithm converges to the proposed estimator.

Furthermore, by executing $R$ local gradient updates per communication, the algorithm limits the total communication complexity to $O(pdK/R)$ per node. In sufficient dimension reduction, the intrinsic structural dimension $d$ is typically a small constant. Therefore, a strategic choice of $R$ achieves communication efficiency, with communication complexity reduced to $O(pK^{1/2})$ . The subsequent section rigorously establishes theoretical guarantees for statistical estimation and algorithmic convergence, explicitly characterizing the precise impact of $R$.

% \section{Theoretical Guarantees}
% In this section, we establish the theoretical guarantees for the proposed method. First, we introduce the regularity conditions required for the subsequent analysis. Then, Theorem~\ref{Theorem:1} provides a non-asymptotic error bound for the estimator of the original problem, followed by a discussion on the adaptivity of the proposed homogeneity pursuit aggregation. Next, Theorem~\ref{Theorem:2} establishes the minimax lower bound for the mean squared error (MSE) of parameter estimation under heterogeneity constraints. This lower bound matches the error upper bound derived in Theorem~\ref{Theorem:1}, which rigorously demonstrates the minimax optimality of our proposed estimator under such constraints. Finally, Theorem~\ref{Theorem:3} analyzes the convergence properties of the solution obtained by Algorithm~\ref{Algo:invexDR} and discusses the influence of the multiple-update parameter $R$ on the communication cost. All the proofs are provided in Appendix \ref{app:proof} of the supplementary material.

% We begin by assuming that the covariates $\x$ and the responses $Y$ are centered, and that the following regularity conditions hold. Specifically, we assume that the data at each node $j$ follow an identifiable regression model, given by $Y_j = m_j(\x_j^\top\bb_j^*) + \epsilon$. Here, $\bb_j^*$ forms a basis for $\calS_{E(Y_j\mid\x_j)}$ and we make general assumptions about this potential true parameter, though basis matrix is not uniquely identifiable.

\section{Theoretical Guarantees} \label{sec:theory}

This section establishes the theoretical guarantees for our proposed framework. Following the necessary regularity conditions, Theorem~\ref{Theorem:1} derives a non-asymptotic error bound demonstrating the adaptivity of our homogeneity pursuit. The model-free prediction error upper bound is also obtained. Theorem~\ref{Theorem:2} establishes the corresponding minimax lower bound, rigorously proving that our estimator achieves minimax optimality under heterogeneity. Finally, Theorem~\ref{Theorem:3} guarantees the convergence of Algorithm~\ref{Algo:invexDR} and characterizes the communication overhead governed by the multiple-update parameter $R$. All technical proofs are deferred to Appendix~\ref{app:proof}.

Assuming centered covariates $\x$ and responses $Y$, the data at node $j$ follow the regression model $Y_j = m_j(\x_j\trans\bb_j^*) + \epsilon$. While the central mean space $\calS_{E(Y_j\mid\x_j)}$ is structurally identifiable, its generating basis matrix $\bb_j^*$ remains fundamentally non-identifiable. The subsequent regularity conditions are formulated with respect to this underlying true parameter.

\vspace{-0.1cm}
\begin{itemize}
  \setlength{\itemsep}{-2pt} % 缩小列表项之间的距离
  \setlength{\parsep}{0pt}   % 缩小同一项内多段落的距离
  \setlength{\parskip}{0pt}

    \item[(C1)] \textbf{The Kernels}: We construct the multivariate kernel as the product of symmetric, univariate kernels. Specifically, the univariate kernel $K(\cdot)$ of order $q$ is required to satisfy
    \[
    \int K(u)du=1,\quad \int u^{i}K(u)du=0,\ 1\leq i\leq q-1,\quad 0\neq\int u^{q}K(u)du<\infty.
    \]
    Furthermore, it has a compact support over which $K(\cdot)$ is Lipschitz continuous.

    \item[(C2)] \textbf{The Density}: For all $j = 1,\dots,m$, both the density function $f(\x_j\trans\bb_j^*)$ evaluated at $(\x_j\trans\bb_j^*)$ and the weight function $w(\x_j) = \{E(\varepsilon^{2}\mid\x_j)\}^{-1}$ are strictly bounded away from zero and infinity.

    \item[(C3)] \textbf{The Smoothness}: Define the conditional expectation $r(\x_j\trans\ba)\defby E\{a(\x_j,Y_j)f(\x_j\trans\ba)\mid\x_j\trans\ba\}$, where the function $a(\x_j,Y)$ can take the form of $Y_j$, $w(\x_j)$, or $\x_j w(\x_j)$. The $(q-1)$th-order derivatives of $r(\x_j\trans\ba)$, $f(\x_j\trans\ba)$, and $m_j(\x_j\trans\ba)$ are  locally Lipschitz continuous in a neighborhood of $(\x_j\trans\bb_j^*)$.

    \item[(C4)] \textbf{The Covariates}: For all $j = 1,\dots,m$, the covariate $\x_j$ follows a sub-Gaussian distribution as same as $\x$. Defining the covariance matrix as $\bSig_j\defby\cov(\x,\x\trans)$, we assume its eigenvalues are bounded; that is, there exists a constant $c>1$ guaranteeing $c^{-1}\leq\gamma_{\min}(\bSig)\leq\gamma_{\max}(\bSig)\leq c$. Additionally, the matrix $\H(\ba)$ must be invertible, when evaluated at any $\bb_j^*$ for $j = 1,\dots,m$, with a constant $\gamma>0$ satisfying $\gamma\I_{pd\times{pd}}\ge \H(\bb_j^*)\geq\gamma^{-1}\I_{pd\times{pd}}$. 

    \item[(C5)] \textbf{The Moment Bound}: The fourth moments are bounded: $E(\norm{\x}_{2}^{4})$, $E(Y^{4})$, and $E\{\norm{\m_{1,j}(\x\trans\bb)}_{2}^{4}\} < \infty$. There exist constants $C_1, C_2 > 0$ such that $E\{\norm{\wh{E}_{j}[\bS\{\x,Y,\ba\}]}_{2}^{4}\} \allowbreak \leq C_1^{4}$ and $E\{\norm{\wh{\H}_j(\ba)-\H(\ba)}_{2}^{4}\}\leq C_2^{4}$. Furthermore, $\wh{\H}_j(\ba)$ is Lipschitz continuous in $\ba$, satisfying $\norm{\wh{\H}_j(\ba_{1})-\wh{\H}_j(\ba_{2})}_{2}\leq L(\x,Y)\norm{\ba_{1}-\ba_{2}}_{2}$, where the bounding function satisfies $E\{L(\x,Y)\} \leq C_3^{4}$. 

    \item[(C6)] \textbf{The Bandwidths}: The bandwidth sequences are chosen to fulfill the asymptotic constraints $Nh_{k}^{2q}h_{l}^{2q}\rightarrow0$, $Nh_{1}^{2(q-1)}h_{l}^{2q}\rightarrow0$, and $Nh_{k}^{d}h_{l}^{d}\rightarrow\infty$ for any index pairs $1\leq k\leq l\leq4$.

    \item[(C7)] \textbf{The Sample Size}: Define the minimum local sample size across all nodes as $n\defby\min_{j=1,\dots,m} n_j$. We stipulate that this sample size satisfies $n\geq\max(c_{1}p,N^{c_{2}})$ for some universally defined constants $c_{1}>0$ and $ 0< c_{2} < 1$.

    \item[(C8)] \textbf{The Similarity}: The similarity across nodes satisfies $\norm{\P(\bb_j^*)-\P(\bb_l^*)}_F \le H$ with $H\ge 0$, for all node pairs $j,l \in \{1,\dots,m\}$, which is equivalent to $\tr\{\P(\bb_j^*)\trans\P(\bb_l^*)\}\leq{1-H/d}$. Besides, there exist $\gamma_1$ and $\gamma_2$ such that $\gamma_2 \le \gamma_{\min}\{(\bb_j^*)\trans\bb_j^*\}\le \gamma_{\max}\{(\bb_j^*)\trans\bb_j^*\} \allowbreak \le \gamma_1 $, for $j = 1,\dots,m$.
\end{itemize}

% \vspace{-0.1cm}
These conditions are mild and commonly adopted in the literature \citep{ma2013semiDRjrssb}. Specifically, condition (C1) allows for the use of second-order kernels, and condition (C6) permits optimal bandwidth selection. Conditions (C2)--(C5) are mild statistical assumptions typically required for nonparametric estimation in sufficient dimension reduction problems \citep{zhu2025distributed}. Condition (C7) is a standard distributed setting assumption \citep{Jordan2019communication}, which allows the local sample size $n_j$ to differ from the global sample size $N$ by a polynomial factor and also bounds the number of nodes $m$ by $O(n^{1/c_2-1})$. Condition~(C8) is introduced in the context of heterogeneous parameter non-identifiability. It utilizes the constant $H$ to quantify both the trace similarity and the signal strength of these parameters. Specifically, a value of $H$ approaching zero signifies a higher degree of parameter homogeneity.
%where $\tr\{\P(\bb_j^*)\trans\P(\bb_l^*)\}\leq{1-H/d}$ as $\norm{\P(\bb_j^*)-\P(\bb_l^*)}_F\leq H$.

\subsection{Nonasymptotic Error Bound of Estimator $\hat\bb$ and Model-Free Prediction}
Next, we present a non-asymptotic error bound for the estimator $\hat\bb$ of the original problem~\eqref{prob:original} as follows.

{\thm [Nonasymptotic Error Bound] Under Conditions (C1)-(C8), if $\lam\le c_3\gamma\gamma_2^{-2}(1+c_2)$ and $H\le c(p/n_j)^{1/2}$ for $j=1,\dots,m$, then it holds that
 \begin{equation}
        \begin{split}
            \norm{\P(\wh\bb_j^{(t)})-\P(\bb_j^*)}_F =&
            O_p\left\{n_j^{-(t+1)/2}+\abs{1-\alpha\lam}n_j^{-1/2} + \lambda{N^{-1/2}}
            +\lambda (H\wedge n_j^{-1/2})\right\}.
        \end{split}
        \label{Theorem1:Result1}
\end{equation}
Furthermore, when $ c_5\gamma^{-1}\gamma_1(1-c_2)\le\lam\le c_3\gamma\gamma_2^{-2}(1+c_2)$ and $t=O(\log{n_j})$ , we have
\begin{equation}
    \norm{\P(\wh\bb_j^{(t)})-\P(\bb_j^*)}_F \le O_p\left( N^{-1/2} + H \land n_j^{-1/2}\right).
    \label{Theorem1:Result2}
\end{equation}
Here $c_3$, $c_4$ and $c_5$ are some generic constants independent of $n_j$, $p$, $d$, $m$ and $H$.
\label{Theorem:1}
}

The non-asymptotic error bound established in Theorem~\ref{Theorem:1} yields three important implications. 

First, the convergence rate of the parameter exhibits a non-monotonic relationship with the regularization parameter $\lam$: it first increases and then decreases. As $\lam$ grows, the amount of information aggregated from other nodes increases accordingly, creating a trade-off between utilizing local sample information and incorporating homogeneous information from other nodes. As $\lam$ continues to increase, the forced aggregation of heterogeneous data introduces substantial bias. This bias, captured by the second and third terms on the right-hand side of \eqref{Theorem1:Result1}, ultimately degrades the convergence rate.   When the regularization parameter $\lam = 0$, the convergence rate in \eqref{Theorem1:Result1} reduces to $ O_p\left( n_j^{-(t+1)/2} + n_j^{-1/2}\right)$, which corresponds to the rate achieved by using only local node information for nonparametric estimation.

Second, with two critical conditions satisfied, appropriate inter-node aggregation and sufficient iterative refinement, we establish the improved convergence rate as result \eqref{Theorem1:Result2} shows. Appropriate aggregation requires carefully tuning the regularization parameter $\lam$ such that $c_5\gamma^{-1}\gamma_1(1-c_2) \le \lam \le c_3\gamma\gamma_2^{-2}(1+c_2)$. Sufficient iteration ensures that the nonparametric elements are estimated with adequate precision. Once the iteration count reaches $t = O(\log n_j)$, the proposed estimator achieves a refined convergence rate of $O_p\bigl(N^{-1/2} + H\land n_j^{-1/2} \bigr)$. This rate is sharper than the convergence rate of local semiparametric estimate $O_p (n_j^{-1/2})$, once the similarity level $H$ satisfies $H \le c_4 pdn_j^{-1/2}$. This restriction for improvement is mild on $H$ and accommodates a broad range of heterogeneity levels.

Third, \eqref{Theorem1:Result2} reveals a two-phase convergence behavior for this refined rate about $m$, representing the distributed network size. For a relatively small network satisfying $m \lesssim (n_j^{1/c_2} \wedge H^2)/n_j$, the convergence rate is dominated by $O_p\bigl( N^{-1/2} \bigr)$. In this initial phase, the error bound strictly decreases as nodes are added. Conversely, for a sufficiently large network where $m \gtrsim (n_j^{1/c_2} \wedge H^2)/n_j$, incorporating additional nodes yields no further reduction in the error bound. In this saturated phase, the rate degenerates to $O_p\bigl( H \wedge n_j^{-1/2} \bigr)$, indicating that further reductions in the error bound require either increasing the local sample size or enhancing parameter similarity. Furthermore, this phase transition highlights that, as the network approaches perfect homogeneity with $H \to 0$, continuously expanding the network with homogeneous nodes can persistently accelerate the convergence rate.

The improved convergence rate of our proposed method stems from its adaptive identification of homogeneous parameter across the nodes. Specifically, the method refines local estimation by dynamically aggregating information from other nodes, leveraging both parameter similarity and the signal-to-noise ratio (SNR) to guide the aggregation process.
To illustrate this, we provide an interpretation for the parametric case with $d=1$. In each local iteration, our method is equivalent to a corrected least squares procedure, whose expression is given as follows.

\begin{align}
        \wh\bb_j&=(\X_{j}\trans\X_{j})^{-1}\X_{j}\trans\y_{j}+(\X_{j}\trans\X_{j})^{-1}\bdelta_j,
        \quad
        \bdelta_j = \sum_{l\ne j } \bdelta_{j,k} 
    \label{eq:corrLS}
    \\ 
    \bdelta_{j,k}  &= ({\lambda}/{m})\left\{ \I_p- \P(\wh\bb_j)\right\}\P(\wh\bb_l)\wh\bb_j/\norm{\wh\bb_j}^2,
    \label{eq:corrLS:term}
\end{align}
where the correction term $\delta_{j,k}$, given in \eqref{eq:corrLS:term}, captures information from other nodes. As illustrated in Figure \ref{Figure:Addptive} in Appendix A of the supplementary material, the magnitude of the correction term depends on both the signal strength of the reference node itself and its similarity to the local node. Therefore, the adaptivity of our method is reflected in two aspects: first, it automatically identifies and aggregates information from nodes with similar parameters; second, it adaptively selects parameters with stronger signal strengths.

Theorem~\ref{Theorem:1} establishes the parameter estimation rate, which subsequent analysis proves to be minimax optimal. However, the final prediction error depends on model-specific hyperparameters, such as the low-dimensional structure dimension $d$ and the underlying true model $\ervm$. To isolate the predictive gains of our proposed method from these confounding factors, we introduce a model-free prediction error metric. Specifically, even without prior knowledge of the exact model structure, we can universally measure the prediction performance using the following mean squared error (MSE) metric:
\begin{equation*}
    E\bigl\{ \norm{\x\trans\P(\wh\bb_j^{(t)})-\x\trans\P(\bb_j^*)}_F^2 \bigr\}.
\end{equation*}
Building upon the estimation guarantees established in Theorem~\ref{Theorem:1}, we directly derive the corresponding non-asymptotic upper bound for this model-free prediction error.

\vspace{-6pt}
{\coro [Nonasymptotic Error Bound for Model-Free Prediction] Under Conditions (C1)--(C8), if $c_5\gamma^{-1}\gamma_1(1-c_2) \le \lam \le c_3\gamma\gamma_2^{-2}(1+c_2)$ and $t=O(\log{n_j})$, we have
\begin{equation}
    E\bigl\{ \norm{\x\trans\P(\wh\bb_j^{(t)})-\x\trans\P(\bb_j^*)}_F^2 \bigr\} \le O_p\left( N^{-1} + H^2 \land n_j^{-1}\right).
    \label{Coro1:Result}
\end{equation}
Here, $c_3$, $c_4$, and $c_5$ are generic constants independent of $n_j$, $p$, $d$, $m$, and $H$.
\label{Coro:1}
}
Crucially, the resulting rate shares the same order as the prediction error established under parametric heterogeneous frameworks \citep{Li2023TLjasa}.

\subsection{Minimax Lower Bound with Heterogeneity Constraints}
We now establish the corresponding minimax lower bound for estimating $\bb$ non-identifiable parameters with the heterogeneity constraint. Following the Assumption (C8), we consider the following heterogeneous parameter space:
\begin{equation}
    \Theta(H) = \Big\{  \bb = \left(\bb_1,\dots,\bb_m \right) \mid
    \bb_j \in\mR^{p\times d}, \, 
    \norm{\P(\bb_j^*)-\P(\bb_l^*)}_F \leq H,\,j=1,\dots,m
    \Big\}.
\end{equation}
The theorem below demonstrates that the proposed estimator is minimax optimal over this parameter space $\Theta(H)$ concerning the MSE  defined as follows,
$$m^{-1}\sum_{j=1}^m\norm{\P(\wh\bb_j)-\P(\bb_j^*)}_F^2.$$
{\thm [Minimax Lower Bound] \label{Theorem:2}
There exists a universal constant $\delta>0$, with the minimum local sample size across all nodes $n$, such that
\begin{equation}
    \inf_{\wh\bb}\sup_{\bb^*\in\Theta(H)}P\left[
        m^{-1}\sum_{j=1}^m
        \norm{\P(\wh\bb_j)-\P(\bb_j^*)}_F^2 \geq O_p\{(mn)^{-1} + (H^2 \land n^{-1})\}
    \right] \geq 1-\delta.
    \label{Theorem:2:result1}
\end{equation}
% With Assumption (C7) and $H \le c_6p/n_j$, we have  % $n_ > c_1p\vee{N^{c_2}}$
% \begin{equation}
%     \inf_{\wh\bb}\sup_{\bb^*\in\Theta(H)}P\left[
%         m^{-1}\sum_{j=1}^m
%         \norm{\P(\wh\bb_j)-\P(\bb_j^*)}_F^2 \geq O_p(N^{-1} + n_j^{-2})\}
%     \right] \geq 1-\delta.
%     \label{Theorem:2:result2}
% \end{equation}
}

Theorem~\ref{Theorem:2} establishes the corresponding minimax lower bound in \eqref{Theorem:2:result1}. Provided that Condition~(C7) holds, the iteration number $t$ is sufficient, and the regularization parameter $\lambda$ is appropriately tuned, Theorem~\ref{Theorem:1} demonstrates that the mean squared error (MSE) bound of our proposed estimator is given by
\begin{equation}
    m^{-1}\sum_{j=1}^m\norm{\P(\wh\bb_j)-\P(\bb_j^*)}_F^2 \le O_p\bigl\{(mn)^{-1} + (H^2 \wedge n^{-1})\bigr\}.
\end{equation}
This upper bound matches the lower bound, thereby achieving the minimax optimal convergence rate over the heterogeneous parameter space $\Theta(H)$. Furthermore, this optimal rate coincides with those established for identifiable parameters in related heterogeneous settings, such as transfer learning \citep[Theorem 2]{Li2022TLjrssb} and multi-task learning \citep[Theorem 2.2]{duan2023MTL}. Consequently, our estimator preserves the same order of minimax optimality even when compared to the strictly identifiable parameter case.
% The corresponding minimax lower bound is established in \eqref{Theorem:2:result1} of Theorem~\ref{Theorem:2}. As demonstrated by Theorem~\ref{Theorem:1},  with the fulfillment of Condition~(C7), a sufficient number of iterations $t$, and an appropriately tuned regularization parameter $\lambda$, the MSE bound of our proposed method reads
% \begin{equation}
%     m^{-1}\sum_{j=1}^m\norm{\P(\wh\bb_j)-\P(\bb_j^*)}_F^2 \le O_p\{(mn)^{-1} + (H^2 \land n^{-1})\}.
% \end{equation}
% which achieves the minimax optimal convergence rate over the heterogeneous parameter space $\Theta(H)$.
% Besides, the order of this lower bound rate also coincides with that established for identifiable parameter, under the heterogeneous setup  of  transfer learning \citep[Thereom 2]{Li2022TLjrssb} and multi-task learning \citep[Thereom 2.2]{duan2023MTL} literature. Consequently, our estimator achieves the same order of minimax optimal rate in the identifiable parameter case
% 给定合适的lam, 可以获得result 1
% 给定合适的, 可以的到result 2

\subsection{Convergence Rate of Algorithm \ref{Algo:invexDR}}

We now establish the convergence properties of the algorithmic sequence $\bb_j^{(t,k)}$ generated by Algorithm~\ref{Algo:invexDR}. Here, $t \in \{1,\dots,T\}$ and $k \in \{1,\dots,K\}$ index the outer and inner iterations, respectively. Conditional on the nonparametric components updated at the $t$-th outer iteration, we define the exact optimal solution to the corresponding inner optimization problem as
\begin{equation}
    \wh\bb^{(t+1)} \in \arg\min_{\ba} G(\ba;\lam,\bb^{(t,0)}).
\end{equation}
Defining the initial inner loop objective gap as $\Delta_0^{(t)} = F_\lam(\bb^{(t,0)};\bb^{(t,0)}) - F_\lam(\wh\bb^{(t+1)};\bb^{(t,0)})$, we present the following convergence rate.

\begin{thm}[Convergence Rate of Algorithm~\ref{Algo:invexDR}] \label{Theorem:3}
Suppose Conditions~(C1)--(C8) hold. Assume the regularization parameter and iteration counts satisfy $c_5\gamma^{-1}\gamma_1(1-c_2) \le \lam \le c_3\gamma\gamma_2^{-2}(1+c_2)$, $n_j^{-1-1/c_2} \le c_4(\lam P)^{-1}$, and $t = O(\log n_j)$ for all $j=1,\dots,m$. Furthermore, let the step size be $\eta = O \{1/(\gamma+\lam\gamma_2)\}$ and the contraction parameter be $\nu = O \{\gamma_2/(\gamma_1+\lam)\}$. Then, the estimation error satisfies
\begin{equation}
    \footnotesize
    \begin{split}
        \norm{\P\{\bb_j^{(t,k)}\} - \P(\bb_j^*)}_F \le O_p\Bigl\{ & (1-\nu)^{k/2}\Delta_0^{(t)}/R + \eta^{k+2}C_1^2 R^2  + n_j^{-(t+1)/2} + N^{-1/2} + (H \wedge n_j^{-1/2}) \Bigr\},
    \end{split}
    \label{Theorem:3:result}
\end{equation}
where $c_3$, $c_4$, and $c_5$ are absolute constants independent of $n_j$, $p$, $d$, $m$, and $P$.
\end{thm}

Theorem~\ref{Theorem:3} establishes in \eqref{Theorem:3:result} that the proposed algorithm converges to the target estimator at the $t$-th outer iteration with a rate of $O_p\bigl\{ (1-\nu)^{k/2}\Delta_0 + \eta^2C_1^2 R^2\bigr\}$. This optimization error bound decomposes into two distinct components. The first term, $O_p\bigl\{ (1-\nu)^{k/2}\Delta_0/R \bigr\}$, captures the linear convergence rate of the projected gradient descent. The second term, $O_p\bigl( \eta^2C_1^2R^2 \bigr)$, quantifies the inevitable drift error introduced by multi-step local updates across the distributed network. These two terms indicate a  trade-off about the multi-step local update parameter, $R$. While larger values of $R$ can significantly accelerate the objective descent in the first term , they concurrently exacerbate the optimization drift, as the second term shows. The domination of the two effect ultimately depends on the specific problem characteristics, particularly the initial gap $\Delta_0^{(t)}$ and the continuity properties $C_1$ from the condition (C5). Consequently, provided a sufficiently large number of inner iterations, specifically $K = O\bigl\{\log(n^{1/2}/R^2)\log^{-1}(1-\nu)\bigr\}$, this optimization error becomes negligible. Under this condition, the algorithmic solution achieves statistical parity with the exact theoretical estimator.

\section{Numerical Simulations}\label{sec:simulation}
We evaluate the proposed InvexDR method via comprehensive simulations, with each configuration replicated 100 times. All computations are executed in MATLAB 2022a on a 16-core, 2.59 GHz workstation with 64 GB RAM. Our code is publicly available at \url{https://github.com/Lear24/InvexDR}. Additional experimental details and ablation studies regarding various constraint configurations are deferred to Appendix~\ref{app:suppResult}.

\subsection{Experimental Settings}
\noindent{\bf Data Generation.} In all scenarios, the response dimension is fixed at $d=2$. The response $Y_j$ at node $j \in \{1, \dots, m\}$ is generated independently from a normal distribution $\mathcal{N}(\ervm(\x_j\trans\bb_j^*), \sigma^2)$ with $\bb_j^* = [\bb_{1,j}^*,\bb_{2,j}^*]$. We consider two distinct configurations for covariates $\x_j \in \mathbb{R}^p$ and the link function $\ervm(\cdot)$:

\noindent\textbf{Example 1.} For each node $j \in \{1, \dots, m\}$, covariates $\x_j \sim \text{Unif}[-2, 2]^{10}$ are generated independently. The response is drawn as $Y_j \sim \calN(\ervm(\x_j\trans\bb_j^*), \sigma^2)$, with the link function $\ervm(\x_j\trans\bb_j^*) = 3(\x_j\trans\bb_{1,j}^*) / \{1 + (1 + \x_j\trans\bb_{2,j}^*)^{2}\}$.

\noindent\textbf{Example 2.} For each node $j \in \{1, \dots, m\}$, covariates are generated as $\x_j \sim \calN(\mathbf{0}, \bSig)$ with $p=16$ and covariance matrix $\bSig = (0.5^{|k - l|})_{p \times p}$. The response follows $Y_j \sim \calN(\ervm(\x_j\trans\bb_j^*), \sigma^2)$, with the link function $\ervm(\x_j\trans\bb_j^*) = \sin(2\x_j\trans\bb_{1,j}^*) \exp(\x_j\trans\bb_{2,j}^*)$.

To systematically induce controlled parameter heterogeneity, we establish a reference coefficient $\bb_1^*$, setting its first four components of $\bb_{1,1}^*$ and $\bb_{2,1}^*$ to $(1, 0, 1, 1)\trans$ and $(0, 1, -1, 1)\trans$, respectively, with the remaining entries zero. For each subsequent node $j \ge 2$, $\bb_j^*$ is generated by rotating $\bb_1^*$ around randomly selected orthogonal basis vectors by an angle $\theta \sim \text{Uniform}\{(0, \theta_{\max}) \cup (\pi - \theta_{\max}, \pi)\}$, where $\theta_{\max} \in (0, \pi/2)$. This rotation ensures the pairwise trace similarity strictly respects the bounds given in Condition (C8) , realistically inducing regulated signal variability while satisfying structural block constraints.

\noindent{\textbf{Baselines and Metrics.}} We benchmark InvexDR against five baselines: local Minimum Average Variance Estimation (MAVE); two local Newton--Raphson approaches under block identity (NR-B) and orthogonal (NR-O) constraints; and their pooled counterparts (pNR-B, pNR-O) to evaluate the empirical gains of cross-node information sharing. Estimation accuracy is evaluated using two standard metrics: the Frobenius norm error $\Vert\mathbf{P}(\widehat{\bb}) -\mathbf{P}({\bb^*})\Vert_F$ and the trace similarity $\tr\big\{\mathbf{P}(\widehat{\bb})\mathbf{P}({\bb}^*)\big\}$.

% \subsection{Performance Evaluation}
% \label{sec:performance}
% In this subsection, we evaluate the performance of our proposed InvexDR method against alternatives. We systematically investigate how key factors, including sample size, number of nodes, distribution of local sample sizes, and noise intensity, affect the estimation accuracy of each method.

\subsection{Performance Evaluation} \label{sec:performance}
We systematically evaluate InvexDR against competing methods across varying sample sizes, network scales, and heterogeneity levels. Detailed node-wise distributions of the $F$-norm and trace similarity metrics are deferred to Figures~\ref{Fig:dF:varn} and \ref{Fig:dTr:varn} in Appendix~\ref{app:suppResult}. 

\begin{figure}[!htbp]
    % \flushleft %左对齐
    \centering
    \subfigure{
        \includegraphics[width =0.3\textwidth]{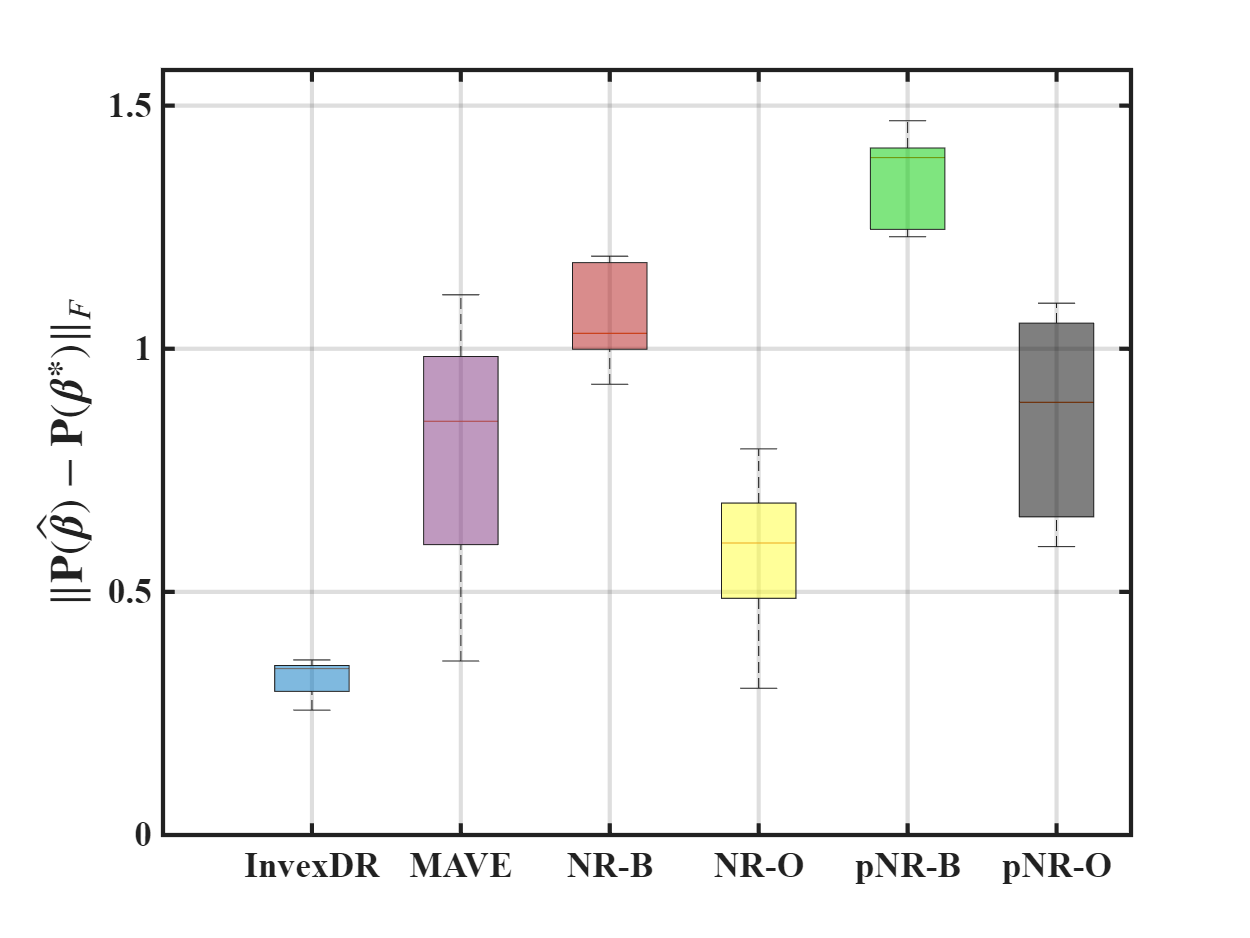}}
    \subfigure{
        \includegraphics[width =0.3\textwidth]{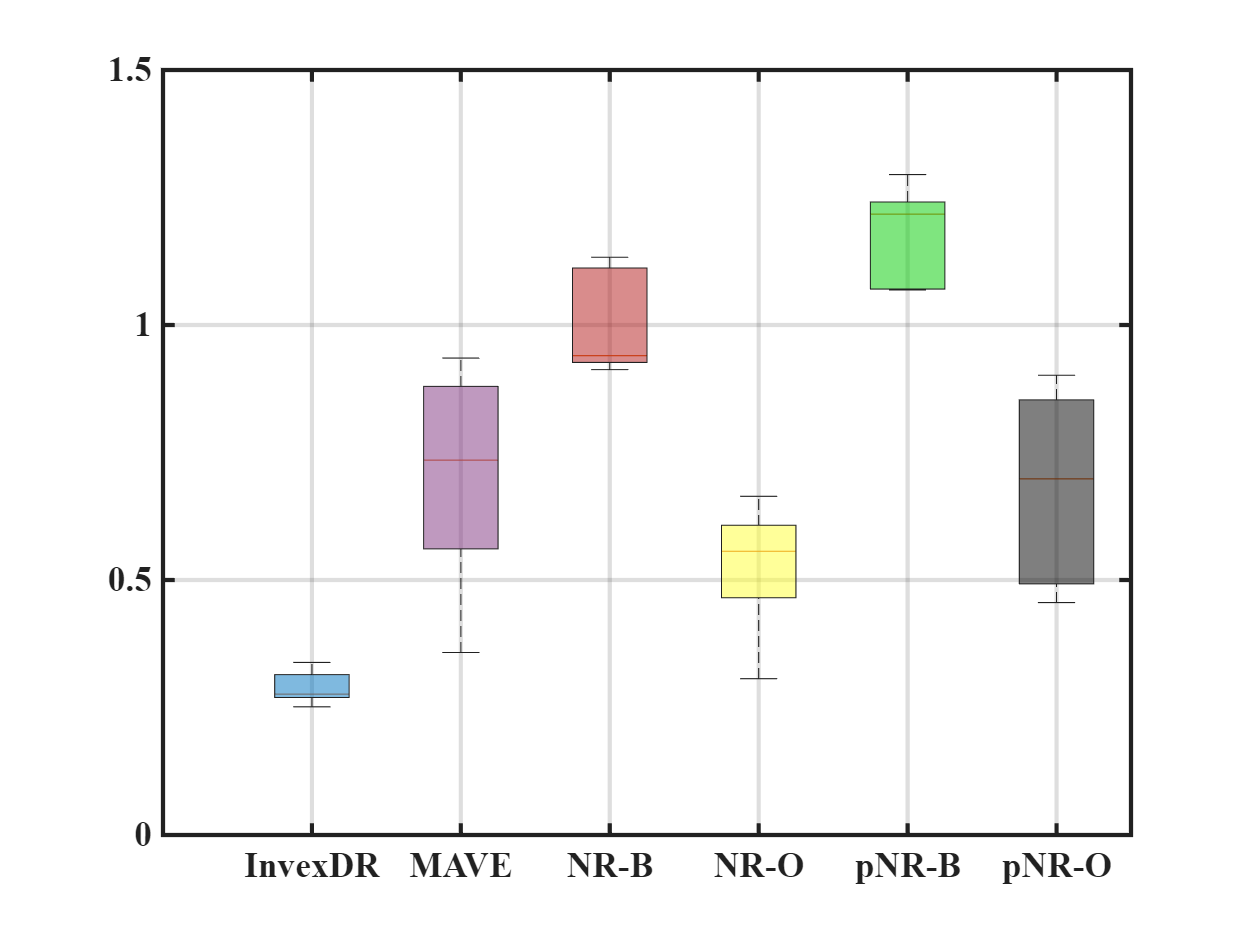}}
    \subfigure{
        \includegraphics[width =0.3\textwidth]{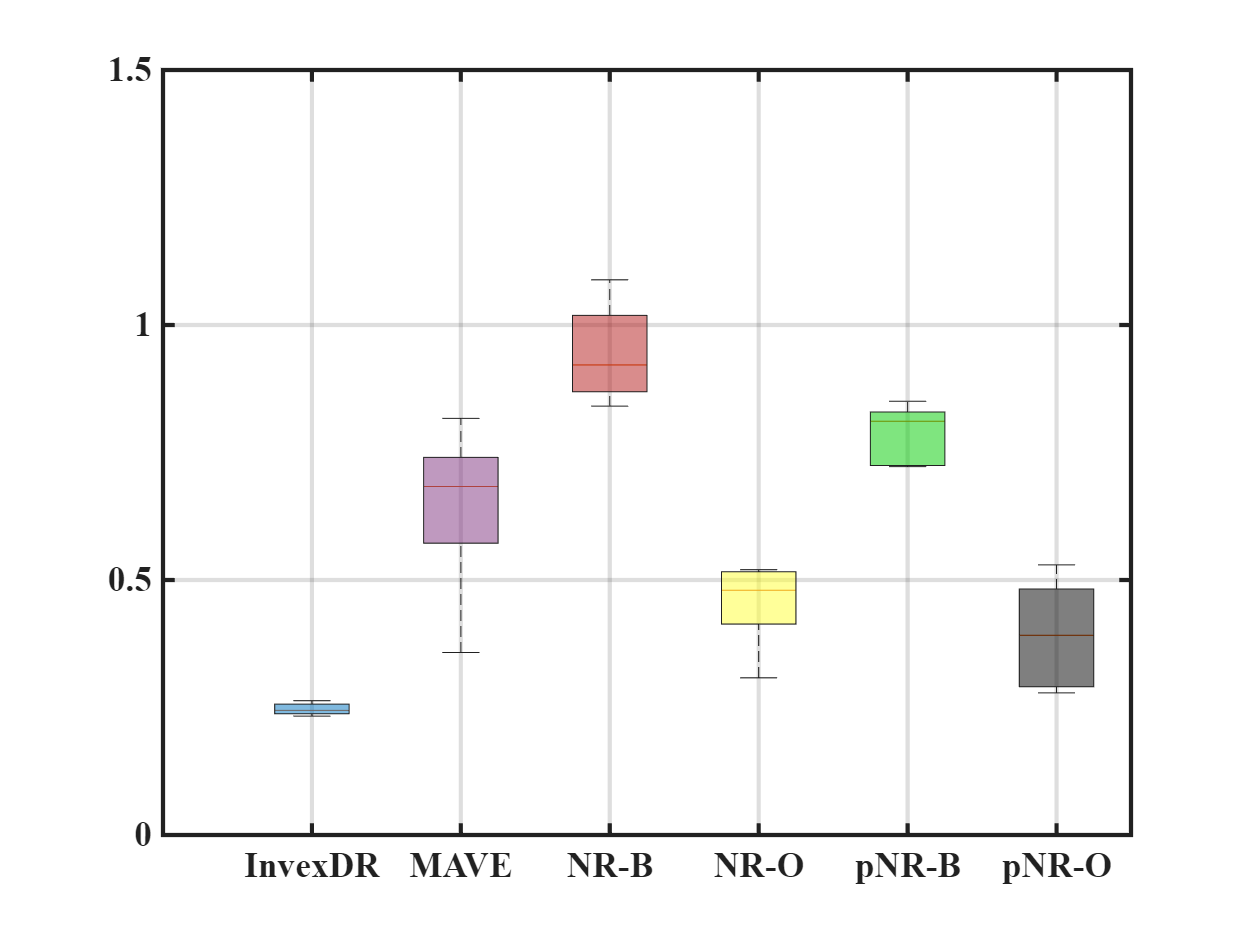}} \\
        \setcounter{subfigure}{0}
        \subfigure[$\theta_{\max}=\pi/3$]{
        \includegraphics[width =0.3\textwidth]{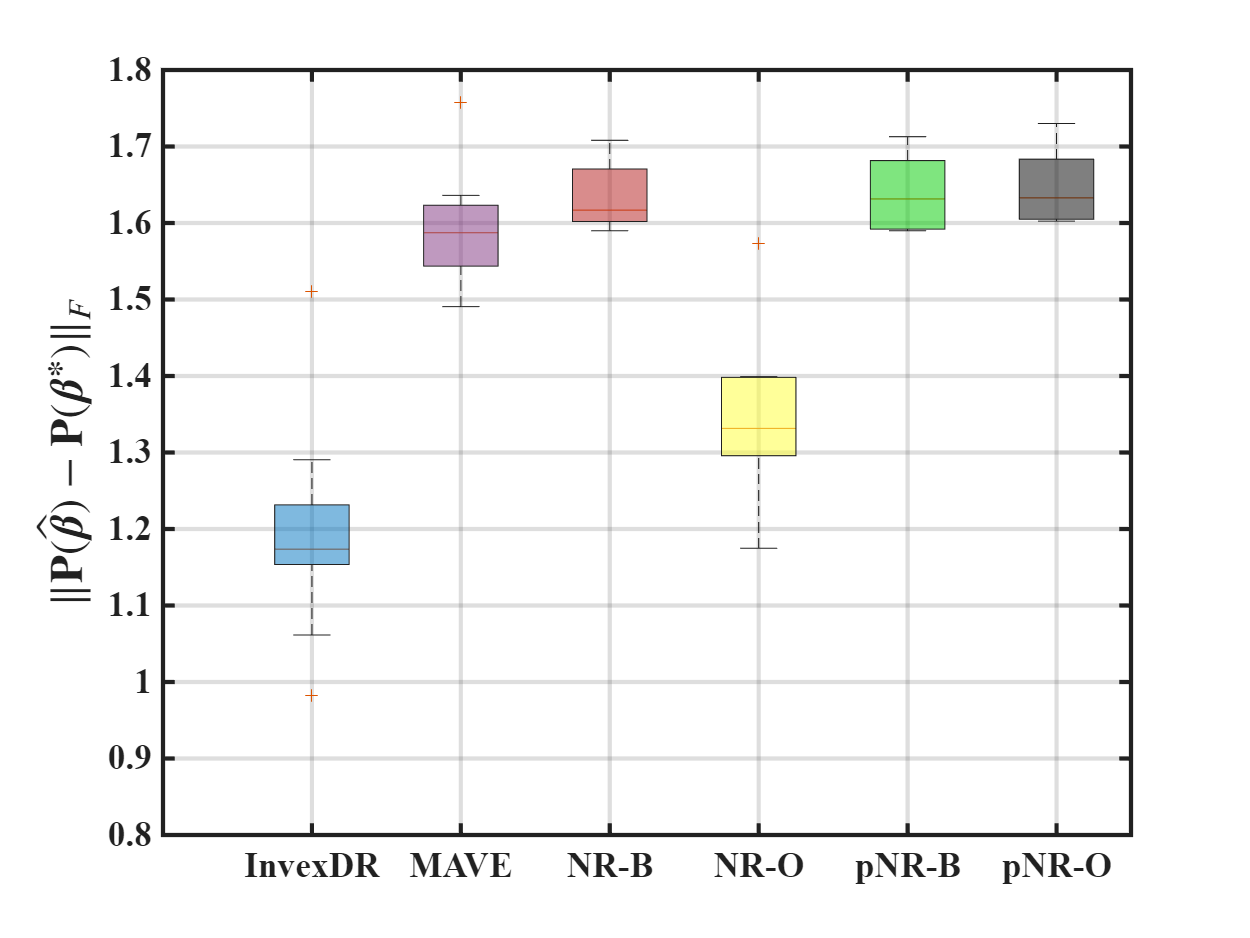}}
    \subfigure[$\theta_{\max}=\pi/4$]{
        \includegraphics[width =0.3\textwidth]{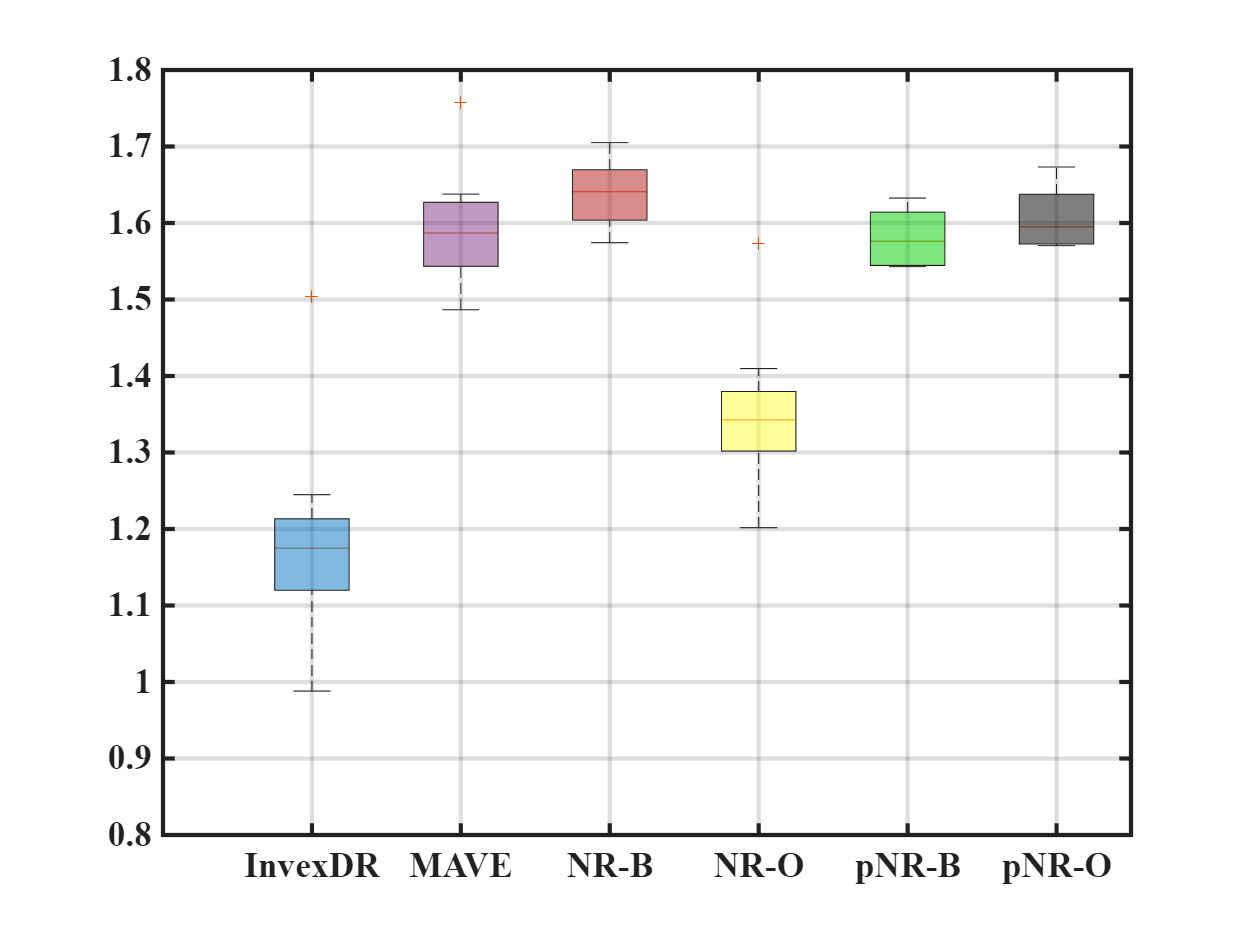}}
    \subfigure[$\theta_{\max}=\pi/8$]{
        \includegraphics[width =0.3\textwidth]{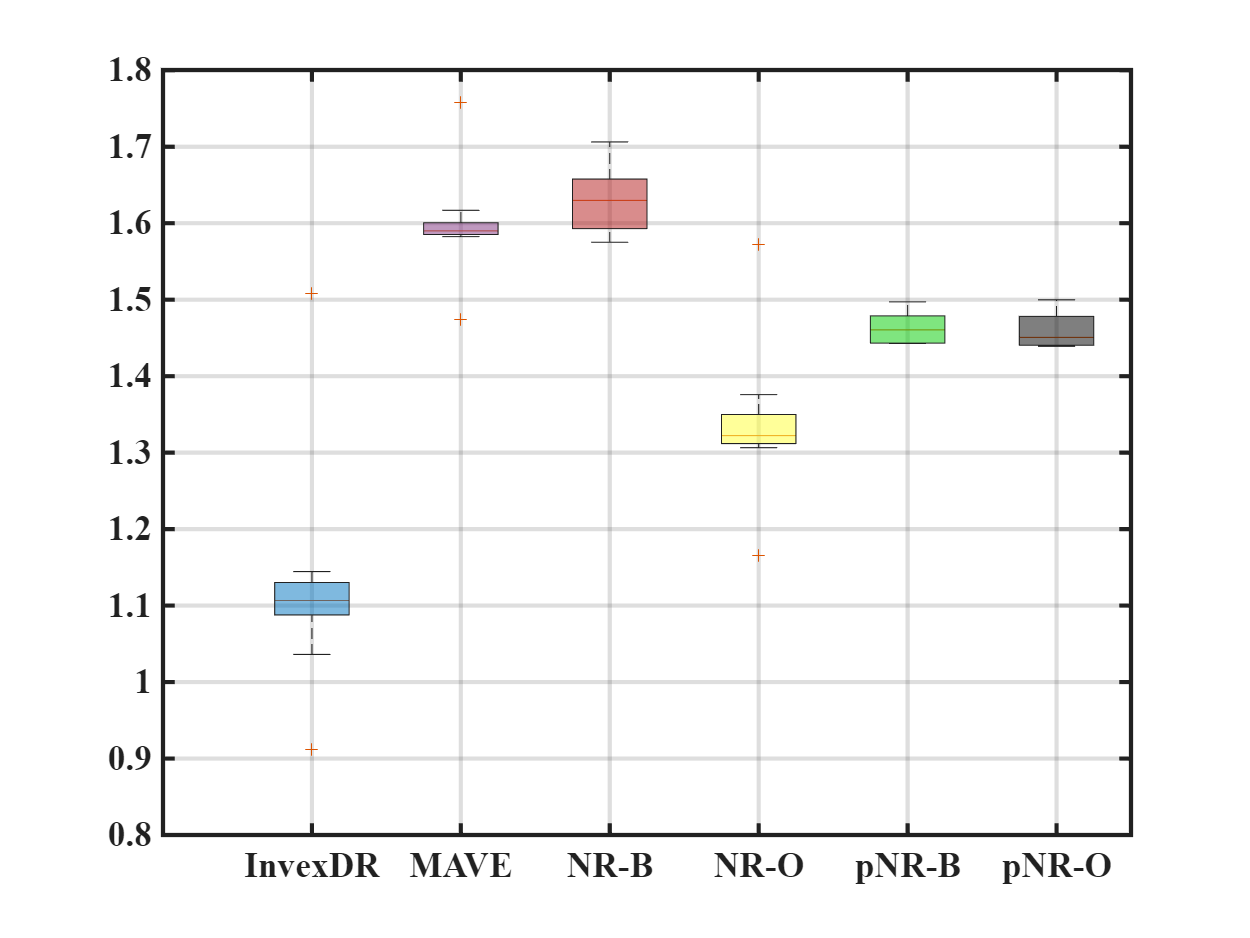}}
   \caption{The $F$-norm error for each node in Example 1 (upper panel) and Example 2 (lower panel) under different similarity levels: (a) $\theta_{\max}=\pi/3$, (b) $\theta_{\max}=\pi/4$, and (c) $\theta_{\max}=\pi/8$. The simulations are performed with $\sigma = 1$ and $n_j=300$ for $j=1,\dots,m$, with $m=5$ for Example 1, and $m=10$ for Example 2.}
    \label{Fig:dF:varn}
\end{figure}

Under a baseline configuration ($\sigma=1, n_j=300$), we evaluate low, medium, and high inter-node similarity ($\theta_{\max} \in \{\pi/3, \pi/4, \pi/8\}$). InvexDR consistently dominates across the entire simulation spectrum. It adaptively leverages latent similarities across the network, yielding proportional performance gains as homogeneity increases. Conversely, isolated estimators (MAVE, NR-B, NR-O) stagnate by ignoring shared information, while pooled estimators (pNR-B, pNR-O) suffer severe degradation even under mild heterogeneity.

\begin{figure}[h]
    % \flushleft %左对齐
    \centering
    \subfigure{
        \includegraphics[width =0.3\textwidth]{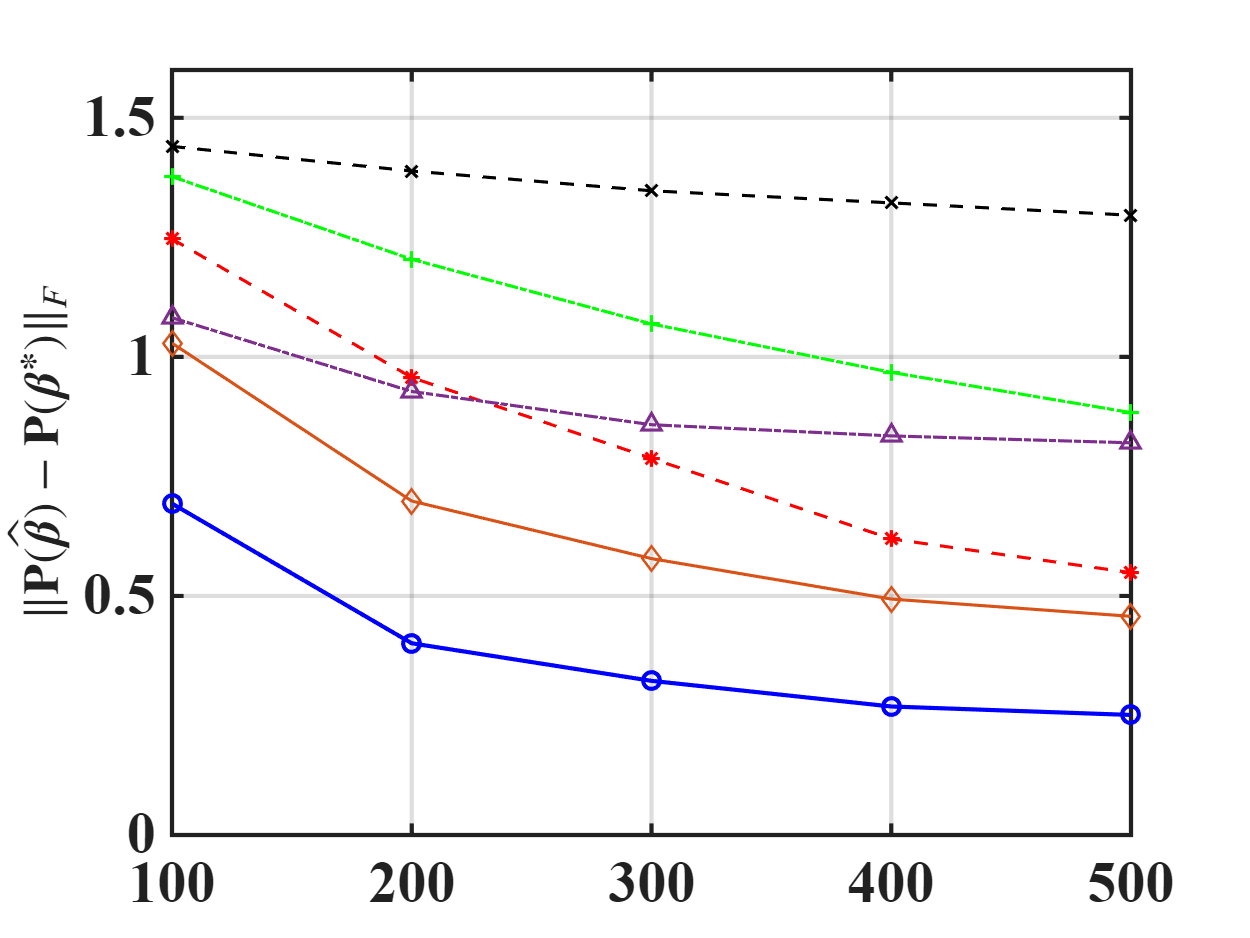}}
    \subfigure{
        \includegraphics[width =0.3\textwidth]{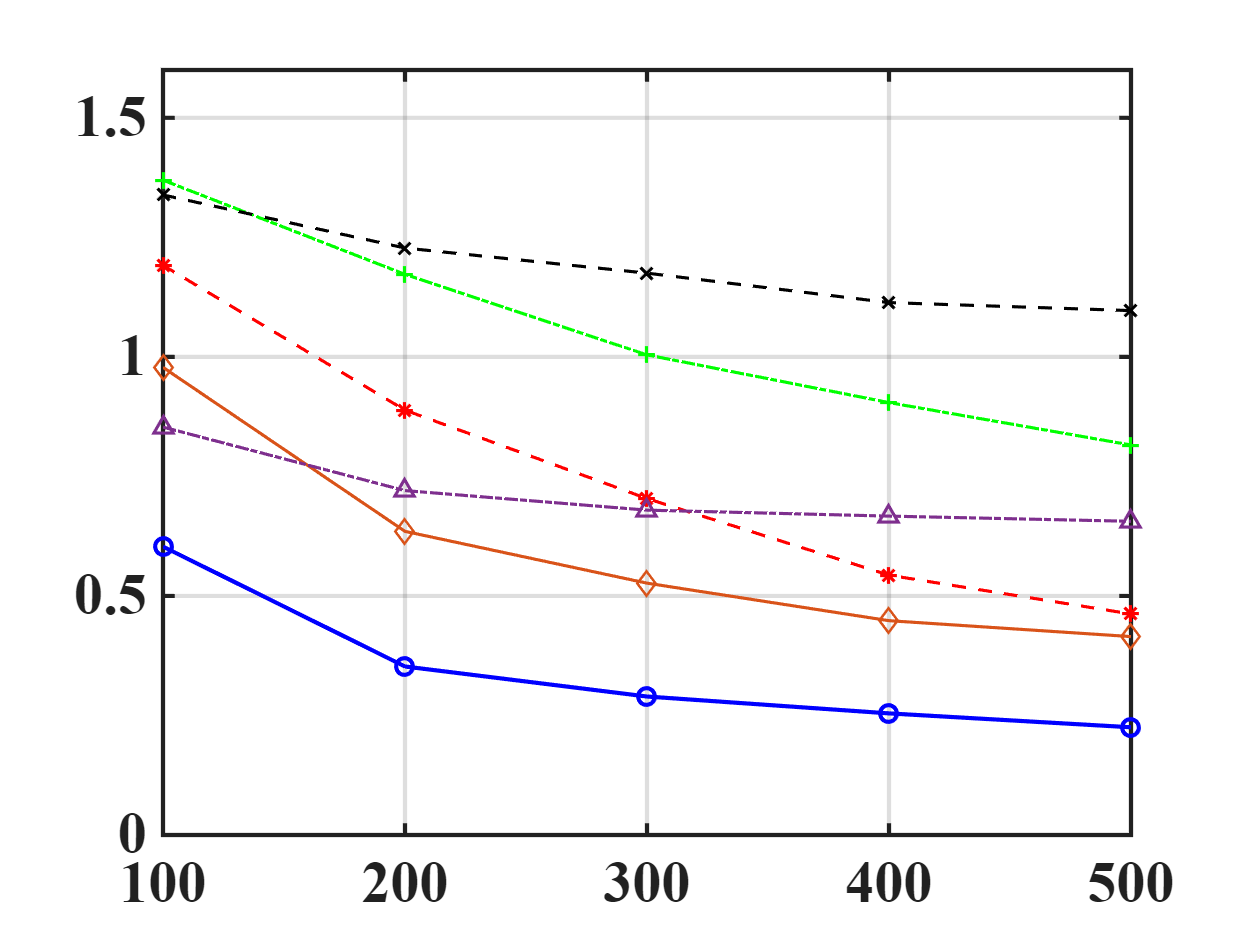}}
    \subfigure{
        \includegraphics[width =0.3\textwidth]{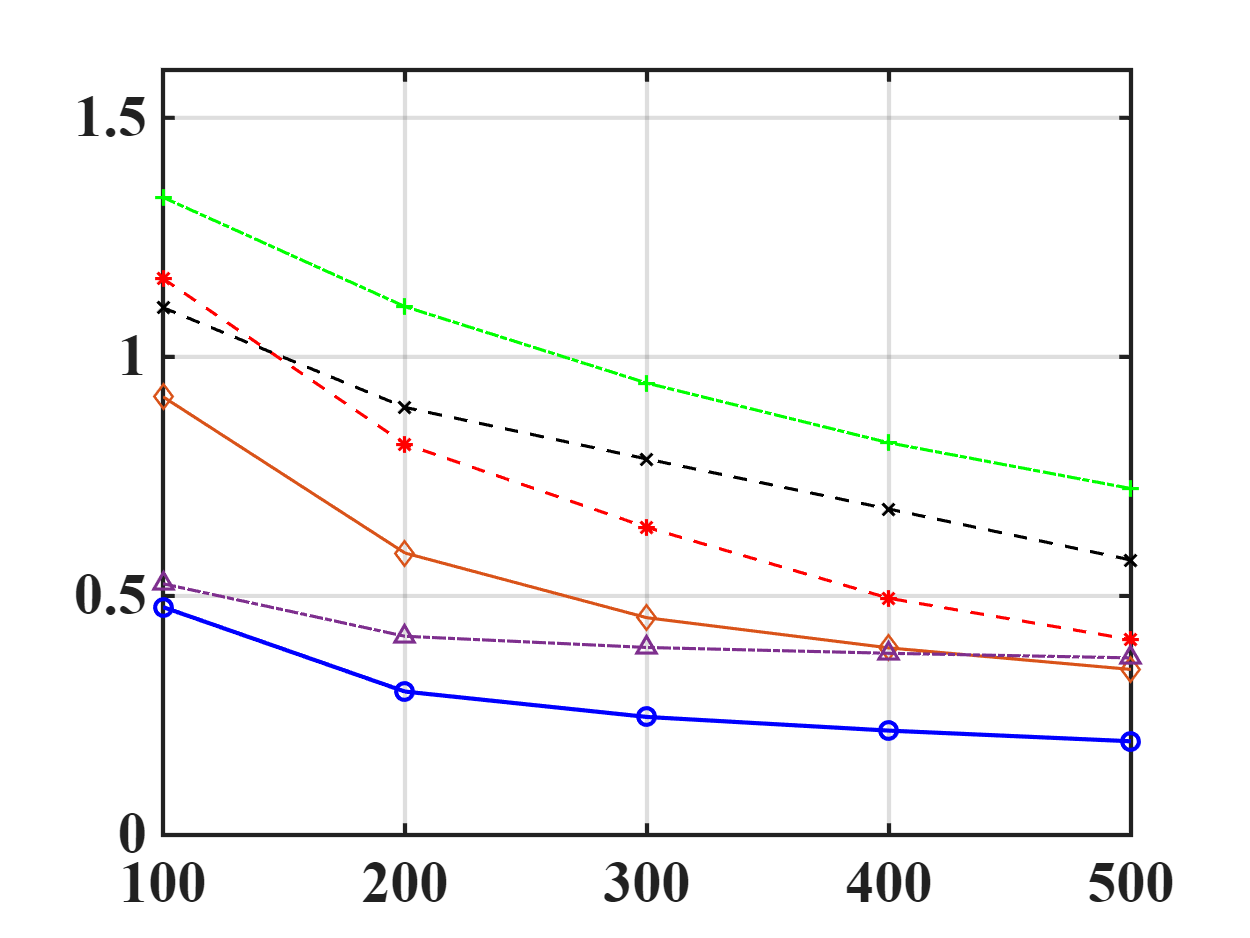}} \\
        \setcounter{subfigure}{0}
    \subfigure[$\theta_{\max}=\pi/3$]{
        \includegraphics[width =0.3\textwidth]{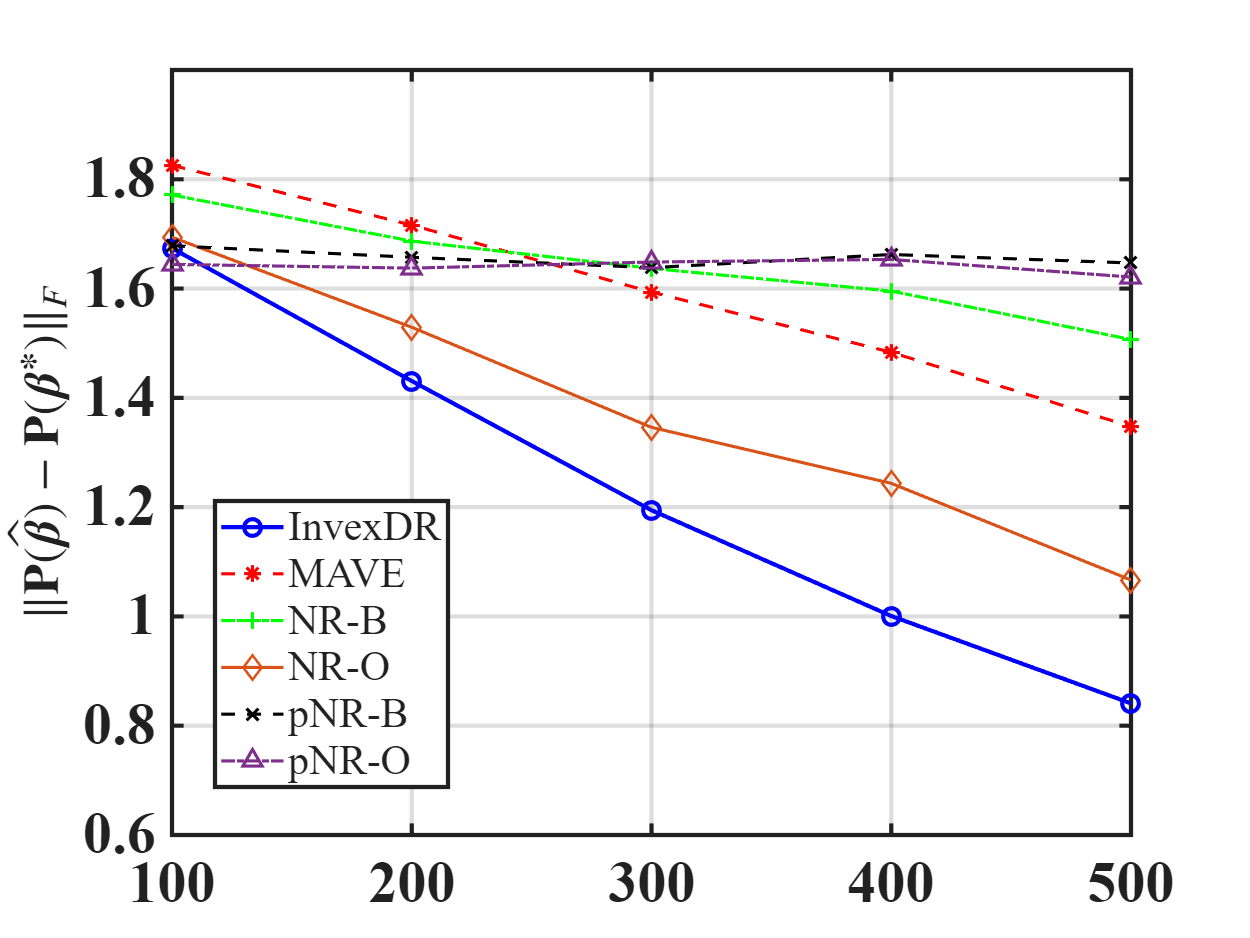}}
    \subfigure[$\theta_{\max}=\pi/4$]{
        \includegraphics[width =0.3\textwidth]{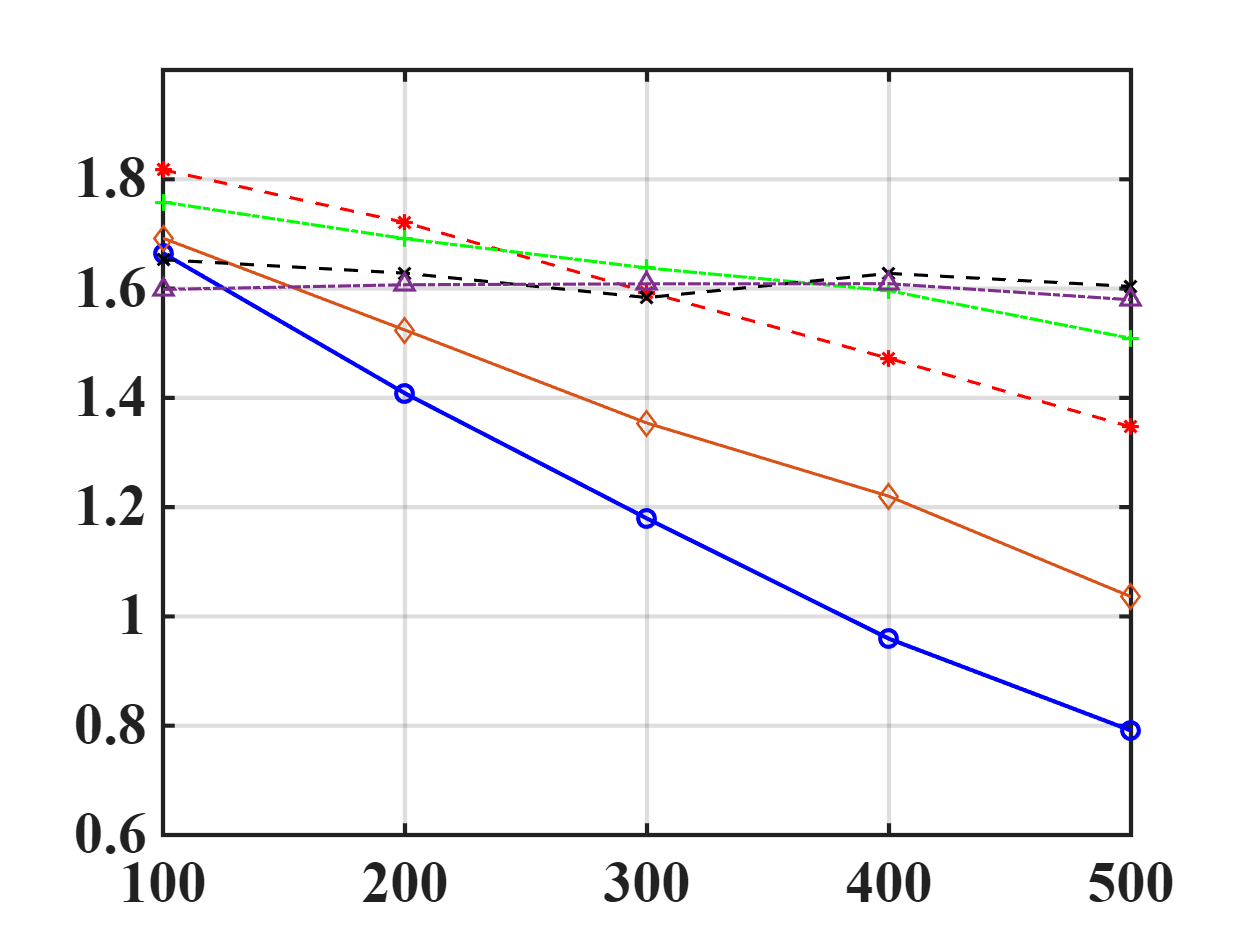}}
    \subfigure[$\theta_{\max}=\pi/8$]{
        \includegraphics[width =0.3\textwidth]{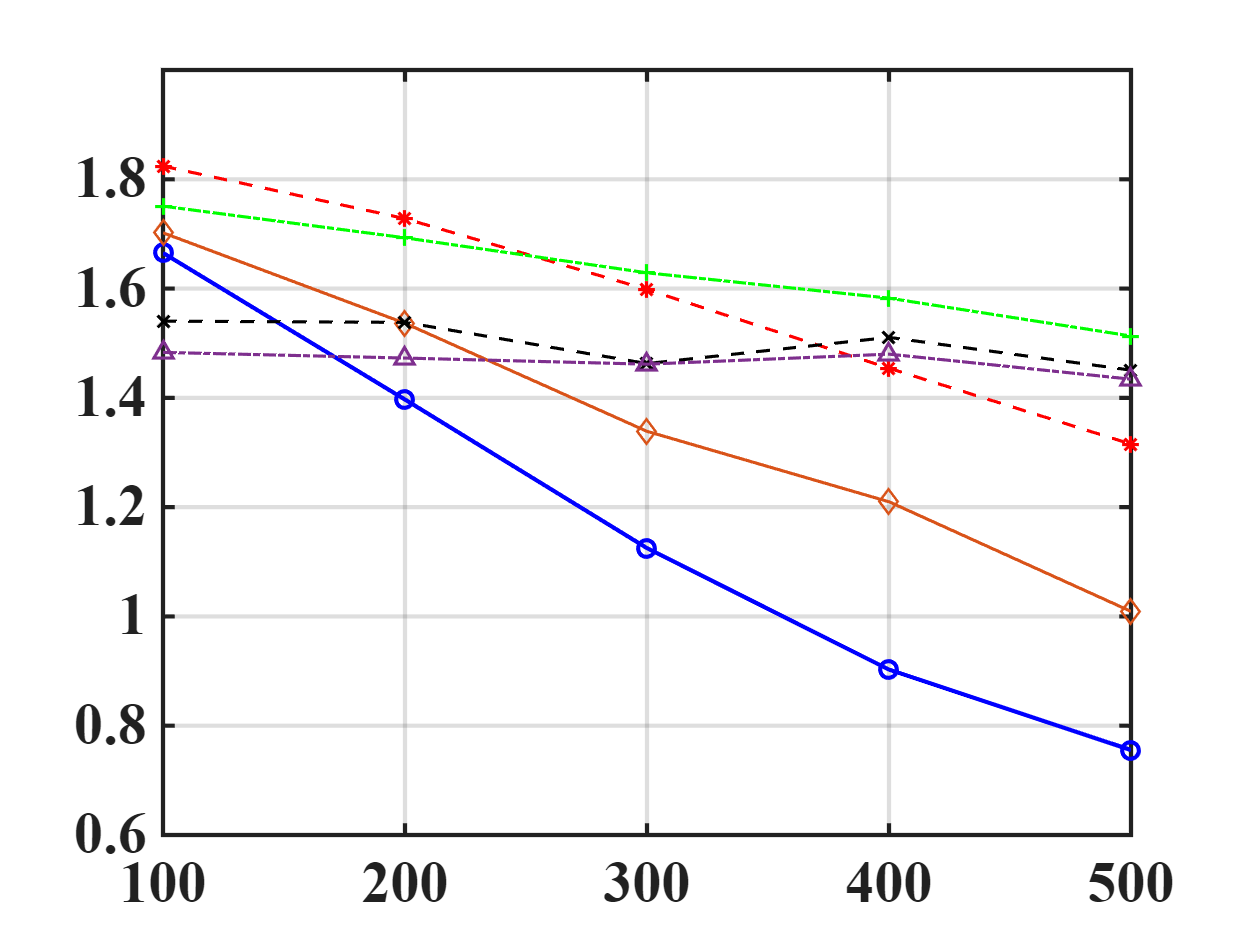}}
    \caption{The average $F$-norm error across nodes is displayed for Example 1 (upper panel) and Example 2 (lower panel) as the local sample size $n$ varies in $\{100, 200, 300, 400, 500\}$, under different angular similarity levels: (a) $\theta_{\max} = \pi/3$, (b) $\theta_{\max} = \pi/4$, and (c) $\theta_{\max} = \pi/8$. The simulations are conducted with $m=5$, $p=10$, $\sigma = 1$ for Example 1, and $m=10$, $p=16$, $\sigma = 1$ for Example 2.}
    \label{Fig:F:varn}
\end{figure}

\noindent\textbf{Sample Sizes.} 
Fixing $\sigma=1$ and varying local measurements $n_j \equiv n \in \{100, \dots, 500\}$, InvexDR exhibits a sharper empirical convergence rate with respect to $n$ (Figures~\ref{Fig:F:varn} and \ref{Fig:Tr:varn}, Appendix~\ref{app:suppResult}). Under heterogeneity, pooled methods fail to improve with larger samples, rendering them strictly inferior to isolated approaches. Notably, because the true parameters inherently violate identity block constraints, the orthogonally-constrained NR-O outperforms NR-B, yet both remain substantially inferior to the unconstrained InvexDR.

\begin{figure}[!htbp]
    % \flushleft %左对齐
    \centering
    \subfigure{
        \includegraphics[width =0.3\textwidth]{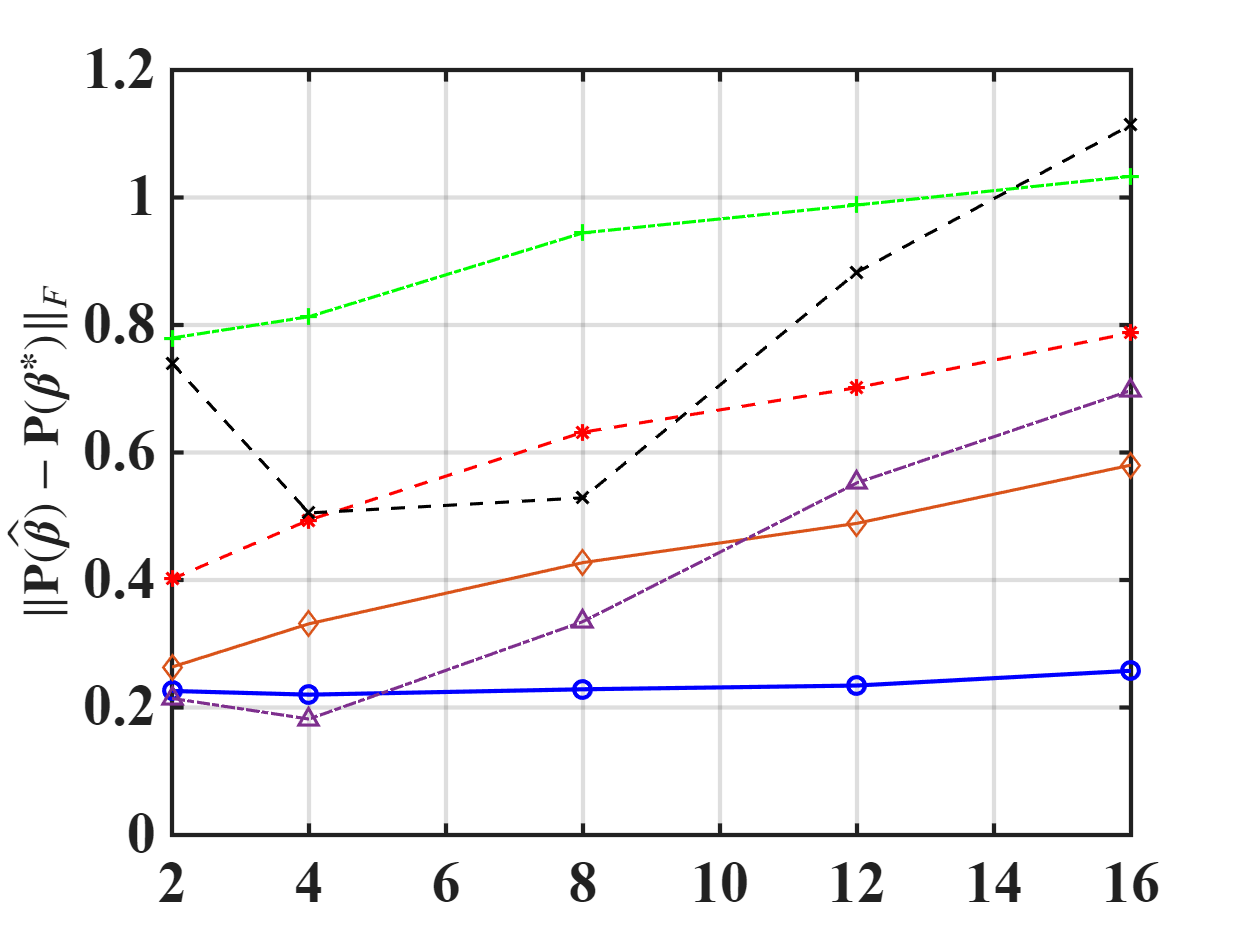}}
    \subfigure{
        \includegraphics[width =0.3\textwidth]{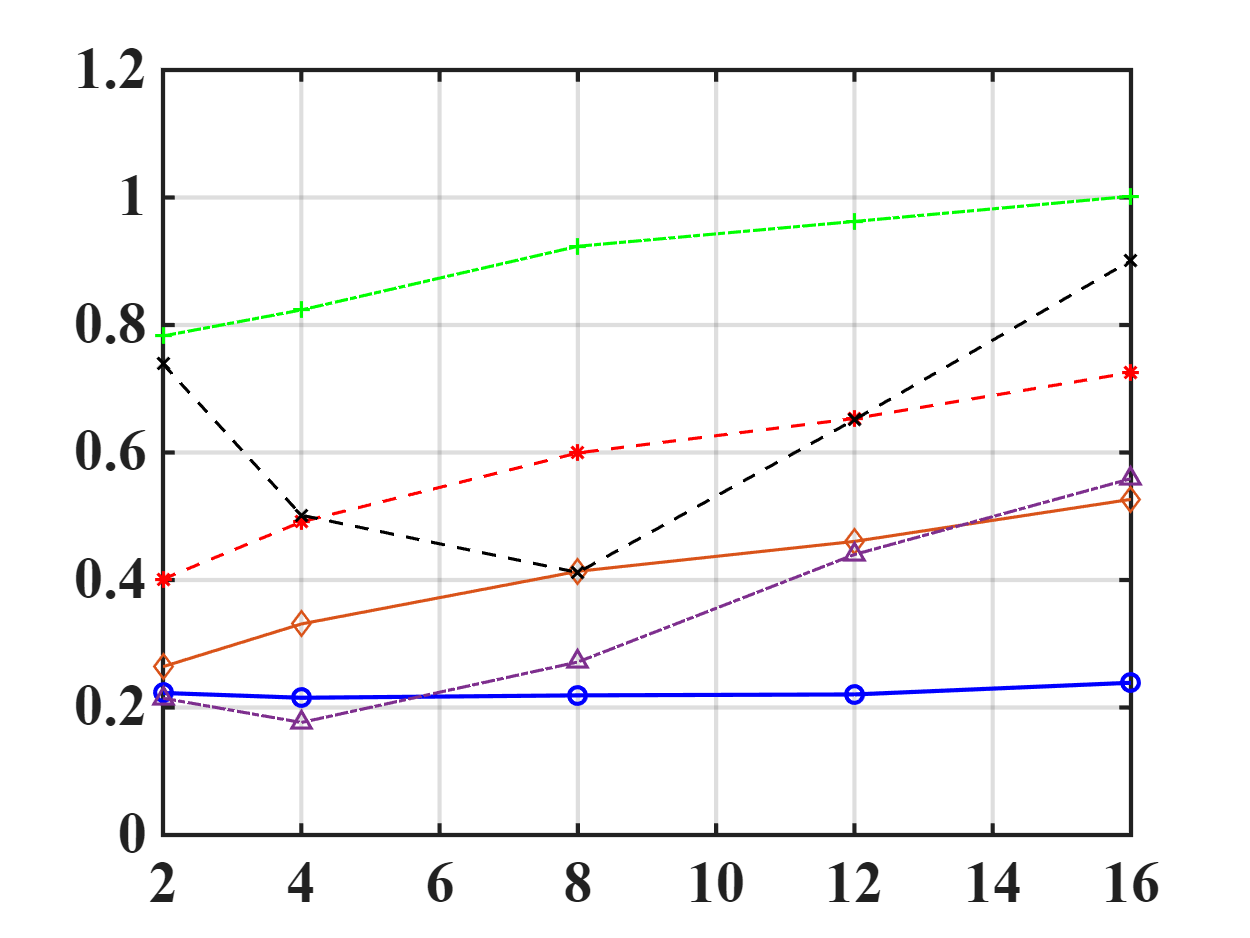}}
    \subfigure{
        \includegraphics[width =0.3\textwidth]{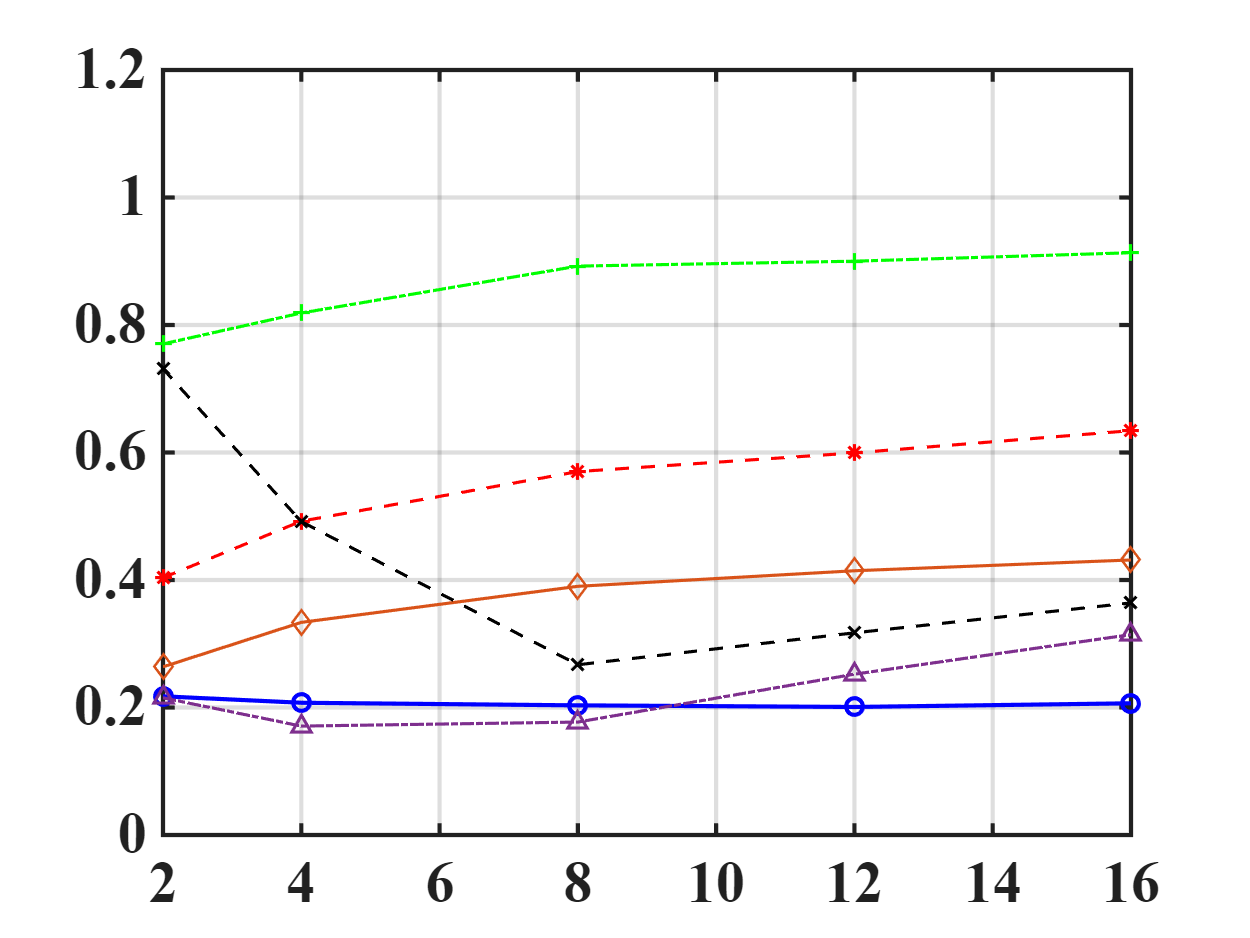}} \\
    \setcounter{subfigure}{0}    
        \subfigure[$\theta_{\max}=\pi/3$]{
        \includegraphics[width =0.3\textwidth]{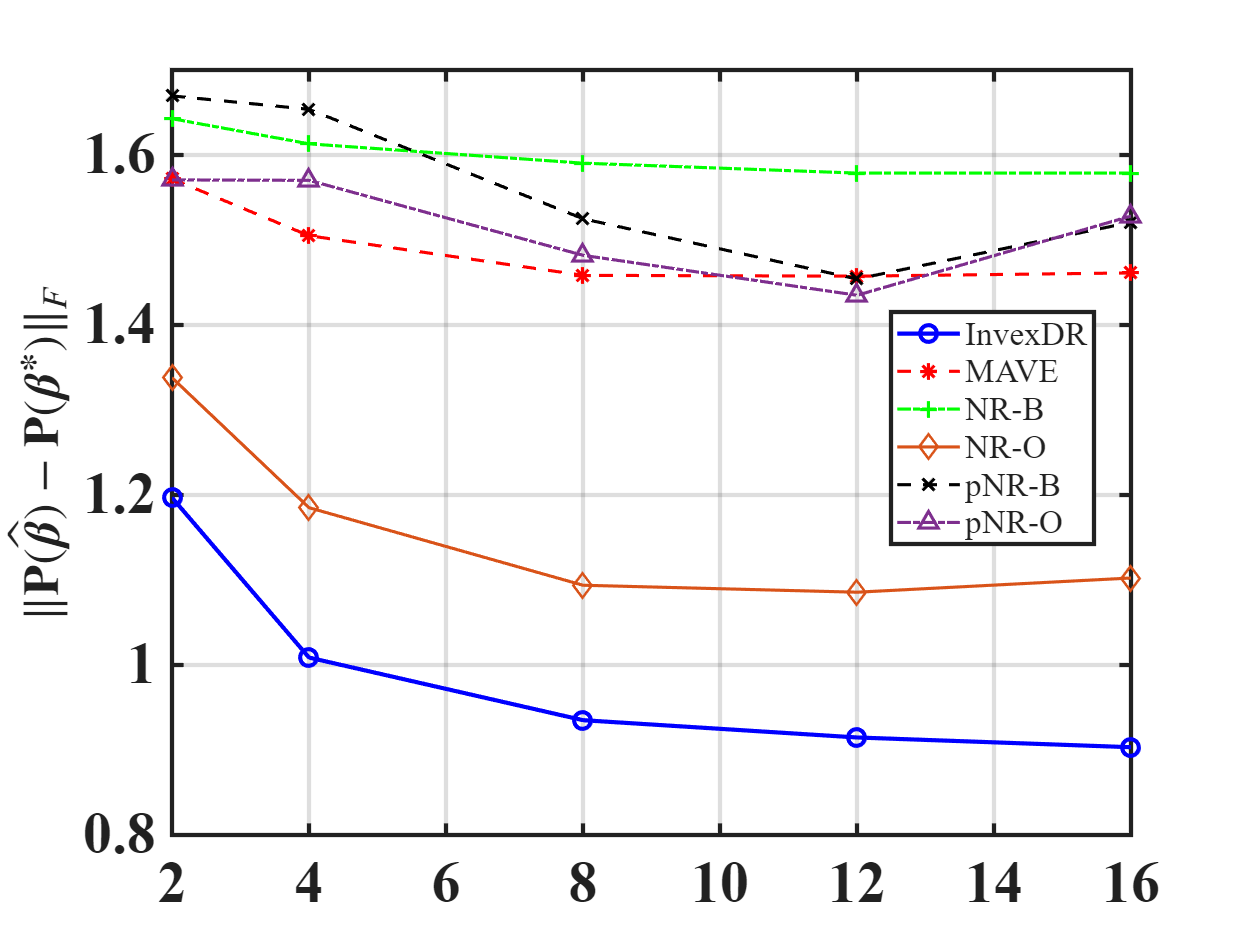}}
    \subfigure[$\theta_{\max}=\pi/4$]{
        \includegraphics[width =0.3\textwidth]{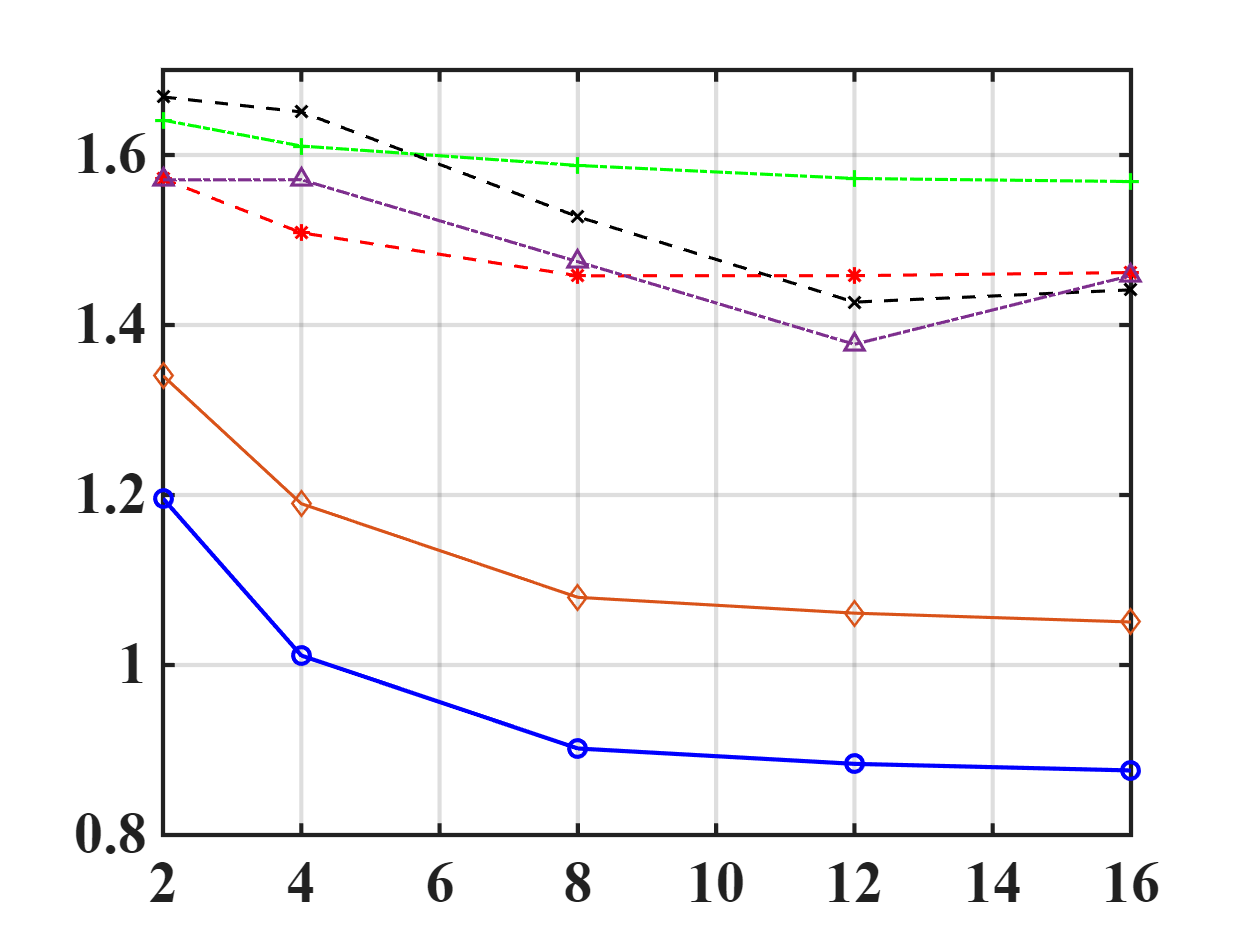}}
    \subfigure[$\theta_{\max}=\pi/8$]{
        \includegraphics[width =0.3\textwidth]{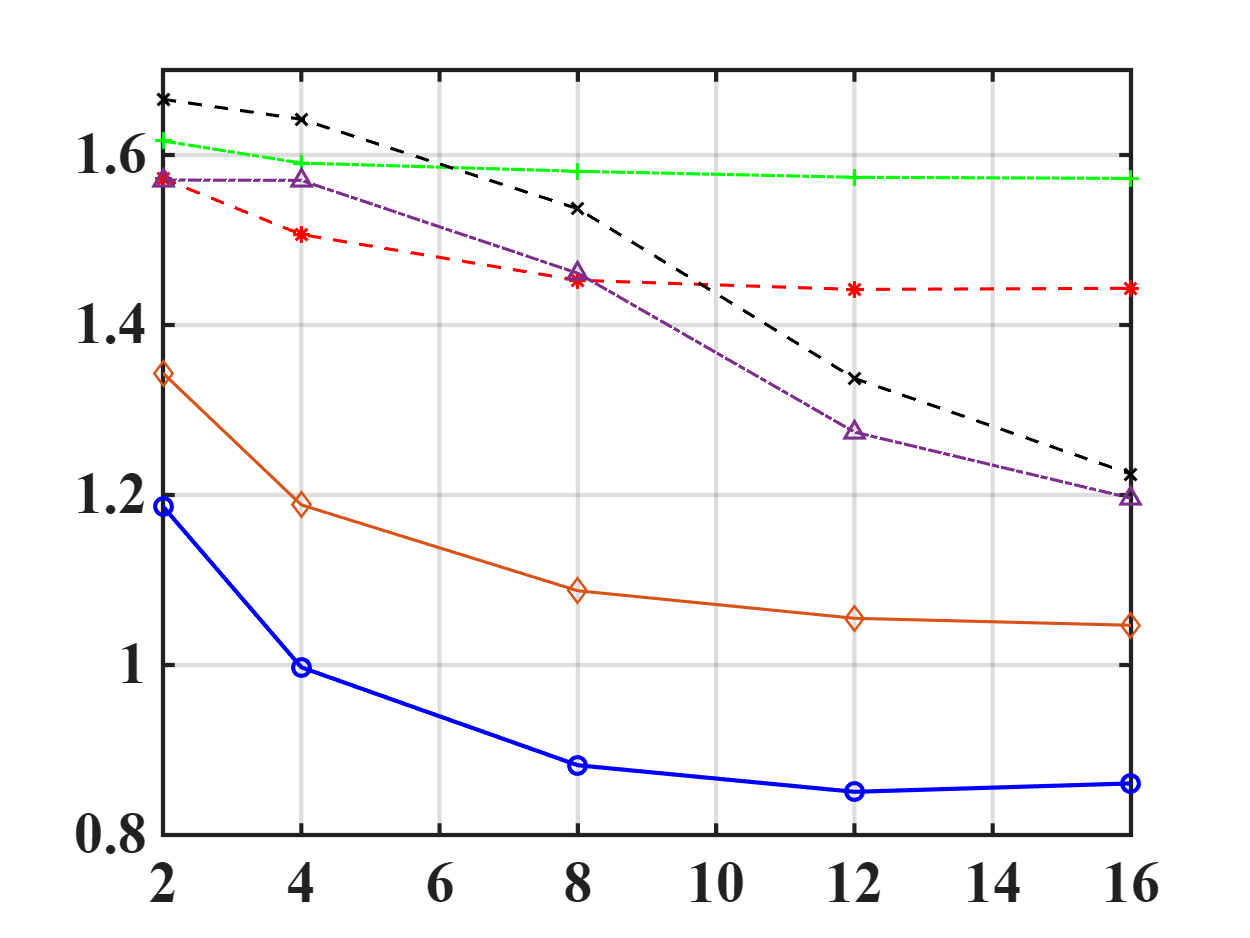}}    
    \caption{The average $F$-norm error across nodes is displayed for Example 1 (upper panel) and Example 2 (lower panel) as the number of node $m$ varies in $\{2,4,8,12,16\}$, under different angular similarity levels: (a) $\theta_{\max} = \pi/3$, (b) $\theta_{\max} = \pi/4$, and (c) $\theta_{\max} = \pi/8$. The simulations are performed with $\sigma = 1$ and $n_j=400$ for $j=1,\dots,m$, where we set $p=10$ for Example 1 and $p=16$ for Example 2.}
    \label{Fig:F:varm}
\end{figure}

\noindent\textbf{Number of Nodes.} 
Fixing $n_j=400$ and varying the number of nodes $m \in \{2, 4, 8, 12, 16\}$, we observe a convergence pattern that perfectly corroborates our theoretical guarantees (Figures~\ref{Fig:F:varm} and \ref{Fig:Tr:varm}, Appendix~\ref{app:suppResult}). Initially, for $m \lesssim (n_j^{1/c_2} \wedge H^2)/n_j$, the error rate is dominated by $O(N^{-1/2})$ where $N = \sum n_j$, improving linearly with $m$. As $m$ grows, the rate transitions to an $O(H \land n_j^{1/2})$ plateau, beyond which additional nodes cease to improve accuracy; lower heterogeneity (smaller $\theta_{\max}$) naturally accelerates this convergence. Furthermore, if $m$ becomes excessively large such that condition (C7) is violated ($n_j \lesssim N^{c_2}$), InvexDR uniquely maintains robust stability without performance deterioration. This sharply contrasts with pooled methods, which actively degrade when absorbing excessive heterogeneous nodes.

% \noindent{\bf Sample Sizes.} To investigate the effect of local sample sizes, experiments were conducted by varying the number of local measurements \(n_j \equiv n \in \{100, 200, 300, 400, 500\}\), with \(\sigma = 1\) fixed. The number of nodes is set to \(m = 5\) for Example 1, and \(m = 10\) for Example 2. The corresponding results are presented in Figure \ref{Fig:F:varn} and Figure \ref{Fig:Tr:varn} in Appendix \ref{app:suppResult}. Compared to other methods, the proposed approach consistently demonstrates superior estimation performance in both low-dimensional and high-dimensional scenarios, effectively adapting to heterogeneity and achieving a sharper convergence rate concerning $n$. In contrast, the pooling methods fail to improve estimation accuracy with increasing sample size when heterogeneity is present, and they are largely ineffective compared to the separate methods under our configuration. However, the proposed method adaptively yields better estimates regardless of the heterogeneity level. Additionally, since the parameters do not satisfy identity block constraints, the NR-B method based on these constraints is outperformed by the NR-O method, which utilizes orthogonal constraints.

The adaptability of our method is reflected in two aspects. First, under varying heterogeneity levels, it can adaptively leverage information from additional nodes to improve estimation performance. Second, even when local estimates are sufficiently accurate or when an excessive number of nodes are introduced, the method remains relatively stable without significant performance degradation.

% distribution of local sample size and Intensity of noise

We also investigate the impact of local sample size distributions and noise intensities; these results are deferred to Appendix~\ref{App:ex:mr}.
Furthermore, we conduct ablation studies to evaluate InvexDR under various constraints and multi-step update schemes. Empirically, our proposed method consistently outperforms constrained alternatives in estimation accuracy. Moreover, the multi-step update strategy achieves estimation precision identical to the single-step baseline, while significantly accelerating algorithmic convergence under appropriate settings. Detailed results are deferred to Appendices~\ref{App:ex:as} and \ref{App:ex:cms} in the Supplementary Material.

\section{Real-Data Application} \label{sec:app}

We apply our proposed method to the eICU Collaborative Research Database \citep{sheikhalishahi2020benchmarking}, which naturally mirrors a distributed framework with inherent heterogeneity. We treat each hospital as an independent node, where strict privacy regulations prohibit direct data pooling.

The clinical task is predicting the Remaining Length of Stay (RLOS) for emergency ICU adult patients presenting with mild-to-moderate coma (Glasgow Coma Scale 9--14). This specific cohort provides high clinical utility for dynamic bed management and treatment evaluation. We extract $p = 14$ clinical variables (3 categorical, 11 continuous), covering demographics, lab measurements, and clinical diagnoses.

To evaluate our method across varying network topologies and local data scarcities, we design two collaborative configurations:

\noindent\textbf{Application 1:} Retaining hospitals with $n_j > 150$, yielding $m = 10$ nodes and a total sample size $N = 2110$ (median $n_j = 192$).

\noindent \textbf{Application 2:} Relaxing the threshold to $n_j > 50$, yielding $m = 26$ nodes and $N = 3499$ (median $n_j = 91$).

We compare our unconstrained InvexDR method against isolated local baselines (MAVE, NR-B, and NR-O) and two constrained distributed variants (InvexDR-B and InvexDR-O). Table~\ref{tab:app:results} summarizes the node-wise mean squared error (MSE) across all clients.

\begin{table}[!htbp]
\centering
\scalebox{0.68}{\begin{threeparttable}	
\caption{The predictive performance of different methods measured by MSE.}
\label{tab:app:results}
\begin{tabular}{@{\extracolsep{12pt}} l cccc c cccc}
\hline \hline
\multirow{2}{*}{Method} & 
\multicolumn{4}{c}{Application 1} & 
& \multicolumn{4}{c}{Application 2}\\
& Mean & Std & Max & Min & 
& Mean & Std & Max & Min   \\
\hline \hline
InvexDR   & \textbf{3.4321} & \textbf{1.3865} & \textbf{5.7710} & 1.7031 & & \textbf{5.3665} & \textbf{3.6224} & \textbf{16.5711} & 0.9973 \\
InvexDR-B & 4.1331 & 1.9552 & 7.7350 & 1.1599 & & 7.5592 & 11.2381 & 59.1050 & 0.8077 \\
InvexDR-O & 5.9153 & 6.0892 & 21.1809 & \textbf{0.5042} & & 9.1009 & 13.4209 & 70.8455 & 1.1945 \\
MAVE      & 5.1133 & 2.7382 & 9.5037 & 1.3417 & & 6.3788 & 4.7499 & 22.3837 & 0.9029 \\
NR-B      & 5.2948 & 3.7485 & 13.9591 & 1.6921 & & 12.1035 & 24.0885 & 99.7232 & 0.6042 \\
NR-O      & 4.3962 & 3.0180 & 12.4054 & 2.1518 & & 6.6344 & 7.4972 & 35.4197 & \textbf{0.3861} \\
\hline \hline
\end{tabular}\end{threeparttable}}
\end{table}

\vspace{-10pt}
Our InvexDR approach consistently dominates all isolated baselines in both average MSE and its standard deviation, achieving at least a $22\%$ improvement in mean predictive accuracy. Notably, despite severe local sample size imbalances, InvexDR ensures uniformly reliable predictions across the network, evidenced by the lowest MSE standard deviation and a substantially reduced worst-case error (max MSE). This confirms that our collaborative training achieves massive synergistic gains, yielding robust network-wide performance.

Interestingly, structural constraints exhibit diverging impacts. While the orthogonal constraint (NR-O) favors isolated learning, the block identity constraint (InvexDR-B) performs better within our distributed framework, suggesting the latent data structure naturally aligns with block identity. Ultimately, by circumventing suboptimal local minima induced by rigid geometric constraints, the unconstrained InvexDR estimator achieves the highest overall predictive accuracy.

\section{Conclusion} \label{sec:conclusion}

This paper develops a communication-efficient distributed framework for estimating non-identifiable parameters under data heterogeneity. By integrating a trace-similarity penalty with an invex relaxation, our method fundamentally resolves severe nonconvexity, guaranteeing stable global convergence from arbitrary initializations. TTheoretically, our estimator achieves a minimax-optimal convergence rate and a tight model-free prediction error bound, matching the fundamental statistical limits established for identifiable parameters. These statistical and algorithmic guarantees are firmly validated by comprehensive simulations and a real-world eICU clinical application.

Future research directions include developing communication-efficient criteria to adaptively select the structural dimension $d$, integrating differential privacy for secure collaborative analysis, and extending the proposed invex framework to complex outcomes, such as right-censored survival data.

% 不同的m如何聚合
% 不同的数据是否可以通过流形的方式进行聚合
\section{Competing interests}
 The authors declare there are no competing interests, financial or otherwise.

\section{Acknowledgments}
Li's work is supported by Qiushi Academic Project of Renmin University of China (NO.RUC25QSDL124).
Sun’s work is supported by National Nature Science Foundation of China (12171479) and the MOE Project of Key Research Institute of Humanities and Social Sciences (NO. 22JJD110001). Zhu’s work is supported by National Key R\&D Program of China (2023YFA1008702) and National Nature Science Foundation of China (12225113 and 12171477).

{
\footnotesize
\centering
\normalem
\bibliographystyle{chicago}
\bibliography{ref}
}

% {
% \footnotesize
% \centering
% \normalem
% \bibliographystyle{abbrvnat}
% \bibliography{ref}
% }

\end{document}